\documentclass[showpacs,nofootinbib,preprintnumbers,prd,superscriptaddress]{revtex4-1}

\usepackage{amsmath}
\usepackage{amsfonts}
\usepackage{amssymb}
\usepackage{graphicx}
\usepackage[titletoc]{appendix}
\usepackage{color}
\usepackage{hyperref}
\usepackage{cleveref}
\usepackage[rightcaption]{sidecap}
\usepackage{subfigure}
\usepackage{comment}
\usepackage{soul}
\usepackage{cancel}

\usepackage{dcolumn}

\usepackage{array}
\usepackage{ctable}
\usepackage{multirow}
\usepackage{siunitx}
\usepackage{longtable}
\usepackage{tabularx}
\usepackage{booktabs}
\usepackage{float}

\graphicspath{{Graphics/}}

\def\be{\begin{equation}}
\def\ee{\end{equation}}
\def\bea{\begin{eqnarray}}
\def\eea{\end{eqnarray}}

\definecolor{vividviolet}{rgb}{0.62, 0.0, 1.0}
\definecolor{amaranth}{rgb}{0.9, 0.17, 0.31}
\definecolor{palatinateblue}{rgb}{0.15, 0.23, 0.89}
\definecolor{brightpink}{rgb}{1.0, 0.0, 0.5}
\definecolor{cornflowerblue}{rgb}{0.39, 0.58, 0.93}
\definecolor{deepcarminepink}{rgb}{0.94, 0.19, 0.22}
\definecolor{radicalred}{rgb}{1.0, 0.21, 0.37}

\hypersetup{ linktoc=all,
    colorlinks, linkcolor={palatinateblue},
    citecolor={brightpink}, urlcolor={amaranth}
}

\begin{document}

\title{Dark energy-matter equivalence  by the evolution of cosmic equation of state}

\author{Anna Chiara Alfano}
\email{annac.alfano@studenti.unina.it}
\affiliation{Dipartimento di Fisica, ``E. Pancini", Universita’ ``Federico II" di Napoli, Via Cinthia, I-80126 Napoli, Italy.}

\author{Carlo Cafaro}
\email{cafaroc@sunypoly.edu}
\affiliation{SUNY Polytechnic Institute,  257 Fuller Road, 12203 Albany, New York, USA.}

\author{Salvatore Capozziello}
\email{capozziello@na.infn.it}
\affiliation{Dipartimento di Fisica, `` E. Pancini", Universita’ ``Federico II" di Napoli, Via Cinthia, I-80126 Napoli, Italy.}
\affiliation{Scuola Superiore Meridionale, Largo S. Marcellino 10, I-80138 Napoli, Italy.}
\affiliation{INFN, Sezione di Napoli, Via Cinthia, I-80126 Napoli, Italy.}

\author{Orlando Luongo}
\email{orlando.luongo@unicam.it}
\affiliation{Universit\`a di Camerino, Divisione di Fisica, Via Madonna delle carceri 9, 62032 Camerino, Italy.}
\affiliation{SUNY Polytechnic Institute, 13502 Utica, New York, USA.}
\affiliation{INFN, Sez. di Perugia, Perugia, 06123, Italy.}
\affiliation{INAF - Osservatorio Astronomico di Brera, Milano, Italy.}
\affiliation{NNLOT, Al-Farabi Kazakh National University, Al-Farabi av. 71, 050040 Almaty, Kazakhstan.}

\date{\today}

\begin{abstract}
We consider a model-independent approach to constrain the equivalence redshift, $z_{eq}$, at which dark energy and the total matter (cold dark matter and baryonic) equate their magnitudes. To this aim, in the context of a homogeneous and isotropic universe, we first consider a generic model where the dark energy contribution is provided by an unknown  function of barotropic fluids. Afterwards, we compute the deceleration and jerk parameters, evaluating at our epoch, namely $z=0$, and at $z=z_{eq}$. Thus, by Taylor expanding around current time the Hubble rate,  luminosity and angular distances, we substitute the theoretical expressions obtained from the aforementioned generic dark energy model, defining a correspondence between quantities evaluated at $z=0$ and $z=z_{eq}$. In so doing, we directly fit these quantities by means of current data sets, involving the most recent Pantheon type Ia supernovae, baryonic acoustic oscillation and Hubble rate points. We consider two hierarchies in our fitting procedures and compare our findings in the spatially-flat universe first and including spatial curvature, later. We assess constraints on the overall equation of state of the universe and its first derivative. We compare our results with those predicted by the standard $\Lambda$CDM paradigm. Specifically, our findings are in agreement at the 2$\sigma$ confidence level, assuming a constant dark energy term. However, our analysis does not rule out the possibility of a slight evolution of dark energy, indicating a small deviation from the scenario of a pure cosmological constant. In particular, the possible departures appear consistent with a phenomenological $\omega$CDM model, rather than more complicated dark energy parameterizations.
\end{abstract}



\maketitle
\tableofcontents

\section{Introduction}\label{introduzione}

Observations in cosmology have confirmed that the universe is currently going through a phase of acceleration, which is attributed to the presence of dark energy \cite{RiessPerlmutter}. This exotic fluid is responsible for the cosmic acceleration by providing a negative equation of state, and its nature is fundamentally different from that of matter and radiation \cite{darkenergyreview}. While we have been able to describe the physical properties exhibited by dark energy, it is still unclear what it represents in terms of constituents, and what kind of microphysics can be associated with it. As a result, various explanations have been proposed in the literature, with the aim of characterizing dark energy based on barotropic fluids \cite{linder, debarotropic}, quantum fields \cite{bagla2003}, modified gravity \cite{MOG1,MOG2,MOG3,MOG4,MOG5,MOG6, dent2013, rubano}, repulsive spacetimes \cite{repulsivegravity1, repulsivegravity2}, symmetry breaking mechanisms \cite{bamba}, and/or other theories (see, for example, \cite{ArmendarizPicon2000, zlatev1998, miao2004, cafaro, cafaro1, cafaro2, cafaro3, colgain}).

Currently, the simplest model used to describe the cosmological background assumes the existence of a cosmological constant, which originates from primordial quantum fluctuations of vacuum energy \cite{coi, carrollambda, sahni1999}. In a way similar to the Casimir effect \cite{casimir}, Einstein's energy-momentum tensor is modified to include a constant contribution from the cosmological constant, which counterbalances gravitational attraction, accelerating \emph{de facto} the universe today. The corresponding paradigm, known as the $\Lambda$CDM model, is based on the existence of two primary fluids: the cosmological constant, denoted as $\Lambda$, and matter in the form of cold dark matter and baryons \cite{peebles2003, carroll, bond, cho2012, ecco,ecco2}. This framework provides an excellent fit to cosmic data, but it is jeopardized by the fine-tuning and coincidence issues, see e.g. \cite{copeland, cdlr}. Additionally, the existence of the cosmological constant is challenged by the so-called \emph{cosmological constant problem} \cite{burgess2013, weilambda}, representing one of the open challenges in modern cosmology\footnote{Some techniques have been proposed to remove the contribution of the cosmological constant by means of cancellation mechanisms. See, for example, Refs. \cite{luongomuccino1, dagostino}, notwithstanding there is no consensus toward these approaches and the problem still persists.} together with the issues related to tensions in cosmological parameters \cite{tension1,tension2,tension3, tension4, tension5}.

In this respect, dark energy is a term used to describe the unknown extension of the $\Lambda$CDM model that includes an evolving equation of state for the component responsible for the acceleration of the universe \cite{hannestad}. However, all dark energy models suffer from the problem of measuring the present matter density and the total equation of state, which are closely related. This leads to a strong degeneracy when different cosmological fluids are measured simultaneously, known as the dark degeneracy problem\footnote{The energy-momentum tensor is postulated to satisfy geometric prescriptions related to Bianchi identities, and Friedmann equations account for the pressure and densities of all components involved in the universe's description. Therefore, it is only possible to measure the total energy density, and the total equation of state is defined as the barotropic factor. This interdependence makes it impossible to draw distinct information on both matter and dark energy, and separate measurements of matter density and the equation of state are insufficient for understanding the physical nature of dark energy.}. This means that it is challenging to establish the properties of dark energy \cite{degenerazione1, degenerazione2, degenerazione3, degenerazione4, kunz}. Dark energy, under whichever form, affects the dynamics of the universe at a given time, \textit{i.e.}, at a precise redshift,  related to the transition epoch \cite{transition1, transition2}. There, the onset of cosmic acceleration occurs. Clearly, it may exist another epoch of the universe evolution in which dark energy and matter, \textit{i.e.}, dark matter and baryons, equate. This period is called equivalence between dark energy and matter, or simply equivalence time (or redshift).  It is important to note that the name ``equivalence between dark energy and matter" is not redundant to avoid confusion with the more well-known equivalence between matter and radiation \cite{galaxybook}.  However, for simplicity, in this paper, we will use the term ``equivalence'' to refer to the equivalence between dark energy and matter.

In order to search for constraints of such periods that do not assume a given dark energy model,  we here propose a \emph{cosmographic} treatment that relates the kinematic coefficients, $q_0$ and $j_0$, the deceleration and jerk parameters at our time, to the redshift at which dark energy and matter become equivalent. Our strategy involves the use of a generic dark energy model, where the fluid responsible for the cosmic speed up is written as a generic normalized function of the redshift. Bearing this in mind, we evaluate the generic deceleration parameter, $q$, and its evolution, $j$ and then we compute these quantities at current time, $z=0$, getting \emph{de facto} $q_0$ and $j_0$, as above stated. Thus, we plug inside the latter terms the mass density, previously rewritten in function of the equivalence redshift. This permits to write up model-independent quantities, namely $q_0$ and $j_0$, in terms of the redshift at which the aforementioned equivalence occurs. We perform this procedure for two main cases, the first involving a spatially-flat universe, the second taking into account a non-vanishing spatial curvature contribution. The effects of spatial curvature are therefore studied, showing how to handle them in view of observations. To do so, we expand around our time the luminosity and angular distances, together with the Hubble rate expansion, and we write them as function of the cosmographic set of parameters, $q_0$ and $j_0$. These resulting expansions are thus \emph{directly} fit with current data, by binning the unknown dark energy function within precise ranges of priors. Consequently, the overall method puts severe constraints at $1\sigma$ and $2\sigma$ confidence levels over the equivalence and free dark energy term for a generic dark energy model. Our findings are thus compared with those obtained in three cosmological models, namely the $\Lambda$CDM concordance background, the $\omega$CDM model and the Chevallier-Polarski-Linder (CPL) parametrization. The equivalence is thus studied between baryons and dark energy, cold dark matter and dark energy and finally between total mass and dark energy. Subsequently, we analyze the implications of our numerical constraints on the equation of state of the universe and its variation. Our analysis reveals that the best-fit background model is consistent with the $\Lambda$CDM paradigm, which is in agreement with previous observational findings. However, we emphasize that a small degree of dark energy evolution cannot be completely ruled out. Therefore, we demonstrate that the $\omega$CDM model can still accommodate our results, suggesting the possibility of a subtle evolution of dark energy throughout the cosmic evolution. Moreover, for our calculations we adopted a system of units with $c=1$.

The paper is structured as follows. In Sect. \ref{sec2}, we highlight the main features of the equivalence redshift and we present its study in a generic dark energy model. In Sect. \ref{sez3}, we derive the expression of the equivalence redshift together with the cosmographic parameters in a flat and a non-flat universe, respectively, for different cosmological models. Moreover, we also derive other equivalence redshift definitions splitting matter into baryons and cold dark matter. In Sect. \ref{sec3}, we describe the statistical analyses used to  constrain the equivalence redshift and the other observables. In Sect. \ref{sezione5}, we show how the equation of state of the universe and its variation evolve in both flat and non-flat universes and, additionally, we discuss the theoretical consequences of our findings. In Sect. \ref{sezione6}, we report our conclusions and perspectives.

\section{Equivalence between dark energy and matter}\label{sec2}

The universe appears homogeneous and isotropic at large-scales. Hence, assuming the validity of the cosmological principle, the maximally-symmetric spacetime that describes the universe dynamics is easily written in terms of the Friedmann-Robertson-Walker (FRW) metric. Including the spatial curvature parameter, $k$, the metric can be recast as 

\begin{eqnarray}
    \displaystyle{ds^2=dt^2-\frac{a(t)^2}{1-kr^2}\left(dr^2+r^2d\Omega^2\right)},
\end{eqnarray}

where $d\Omega^2\equiv d\theta^2+\sin^2{\theta}d\phi^2$. 

Plugging this metric into the Einstein's field equations, providing the well-known Friedmann equations expressed by

\begin{equation}\label{Fried}
H^2 \equiv \left(\frac{\dot{a}}{a}\right)^2 = \frac{8\pi G}{3}(\rho+\rho_k)\,,\quad
\dot{H}+H^2  = -\frac{4\pi
G}{3}\left(\rho+3P\right)=-\frac{4\pi
G}{3}\left(1+3\omega\right)\rho\,,
\end{equation}
Here, $H$ denotes the Hubble parameter, $H\equiv\frac{\dot a}{a}$, $P$ the total pressure associated with the fluid, whose density is $\rho$ while $\rho_k\equiv -k\,a^{-2}$ is the spatial curvature density.

Moreover, we stress that, regarding the pressure $P$ and density $\rho$, since more than one species can enter the energy-momentum tensor, denoting with the subscript ``i" a single species, we easily have $\rho=\sum_i\rho_i$ and $P=\sum_i P_i$, implying  $\omega\neq\sum_i\omega_i$. Hence, to specify a given cosmological model, we need to invoke the equation of state of the net cosmic fluid, which is denoted by $P= P(\rho)$ or $\omega=\frac{P}{\rho}$ while in order to
characterize the single fluid one has to know its specific equation of state, namely $\omega_i\equiv\frac{P_i}{\rho_i}$.

Unfortunately, the information regarding each barotropic factor associated with every single subcomponent cannot be obtained from the \emph{total} equation of state, leading to a problem of cosmic degeneracy \cite{degenerazioneultima}. This issue limits our understanding of the overall dynamics as it is not easy to constrain every single fluid magnitude from observations. The issue arises due to the fact that mass degenerates with dark energy in the total energy density of the universe. To partially bypass this problem, we can focus our approach on late-times and arbitrarily split $\omega$ into (at least) three parts: matter, dark energy, and spatial curvature, while neglecting radiation as well as other exotic components such as magnetic monopoles and topological defects, which do not have a significant impact on the large-scale dynamics of the universe today.

In this scenario, the need for a model-independent technique of cosmological reconstructions becomes particularly relevant. This is required to quantify the kinematics and dynamics of the universe, which is viewed as a whole dynamical system, whose energy-momentum tensor is not fully-known \emph{a priori}, owing to the existence of dark energy.

Among all the model-independent treatments, Taylor expanding the scale factor $a(t)$ can provide valuable information on the derivatives of the Hubble rate \cite{e1,e2,e3}. By substituting the expanded scale factor into observational quantities, we can gain insights into the dynamics of the universe.

To show this fact, let us write the scale factor expansion around our time, namely $t_0$, corresponding to late-times, or alternatively small redshift, say $z=0$, by\footnote{Conventionally, $a(t_0)=a_0=1$, as commonly assumed. }

\begin{eqnarray}\label{serie1}
a(t)= 1 + H_0 \Delta t - {1\over2} q_0 H_0^2 \Delta t ^2 +{1\over6} j_0 H_0^3\Delta t ^3 + \ldots\,,
\end{eqnarray}

where $\Delta t\equiv t-t_0>0$. Denominating the deceleration parameter, at any time, $q(t)$, and the jerk parameter, associated with the $q(t)$ variation, $j(t)$, we can write

\begin{eqnarray}\label{q_j_definition}
q(t)&=&-\frac{\dot{H}}{H^2} -1\,,\\
j(t)&=&\frac{\ddot{H}}{H^3}-3q-2,
\end{eqnarray}

where we have written the deceleration and jerk parameter in terms of the Hubble rate. The subscript ``0", reported in the quantities $q_0$ and $j_0$ in Eq. \eqref{serie1} indicates that the relations \eqref{q_j_definition} are computed at $t=t_0$. It is well-established that $q$ indicates whether the universe accelerates, as $q_0<0$, or decelerates, as $q_0>0$, and $j$, if positive, denotes whether the universe changes its acceleration sign after the transition time \cite{transition1,transition2}.

Constraints over these quantities have been tightly found throughout recent decades \cite{constraints1, constraints2, luongoprd2012}, showing plausible priors, being valid for any cosmological model under exam, \textit{i.e.}, for any form of dark energy evolution. These constraints may be naively summarized as \cite{constrq, mamon1, mamon2}

\begin{align}
|q|&\in[0.5;1]\,,\\
j&\in[0.5;1.5]\,,
\end{align}
that may be used as direct priors to employ during fitting procedures. Particularly, exceeding the above bounds would indicate that the cosmological model under exam fails to be predictive at all stages of the universe evolution. For instance, a model that shows a negative jerk parameter today is clearly ruled out \cite{bravetti}. Conversely, fulfilling the aforementioned limits appears as a necessary but not sufficient requirement to guarantee the goodness of a given cosmological model, since the technique of Taylor expansion, prompted in Eq. \eqref{serie1}, turns out to be valid only at  background cosmography. Indeed, the domain of small perturbations, or more precisely early-stage epochs are not included into this picture.

Nevertheless, using Eq. \eqref{serie1} in the cosmic distance calculation offers a significant advantage by providing expanded versions of these quantities that can be directly compared with data, without needing the explicit form of the dark energy term. This technique is known as cosmography or cosmokinetics \cite{visser, visser2004, cattoen2008, capozziellobenetti}, and has been studied in various contexts, even beyond general relativity, as seen in \cite{constraints2}.

We will delve further into this point in the upcoming sections. For now, we will focus on how to relate $q$ and $j$ in terms of a generic dark energy model, to establish limits on the dark energy abundance without fixing it. In the next subsection, we will outline our strategy, where the only assumption is that the functional forms of density and pressure in Eqs. \eqref{Fried} remain unchanged. We will then naively assume that dark energy evolves as a barotropic perfect fluid, expressed in terms of the redshift $z$.

\subsection{A generic dark energy model}

To develop a generic dark energy term, built up in terms of an unknown function, say $G(z)$, driving the universe to accelerate, one can assume that the first Friedmann equation is under the form \cite{malcap}
\begin{equation}\label{hz}
H(z)=H_0\sqrt{\Omega_{m}(1+z)^{3}+\Omega_k(1+z)^2+\Omega_{DE}G(z)}\,,
\end{equation}
where the normalized densities are defined as $\Omega_{m}\equiv\rho_m/\rho_c$,  $\Omega_{k}\equiv\rho_k/\rho_c$ and $\Omega_{DE}\equiv\rho_{DE}/\rho_c$, with $\rho_c\equiv\frac{3H_0^2}{8\pi G}$ the current value of the critical density.

Clearly, the dark energy density, $\Omega_{DE}$ or $\rho_{DE}$ appears unspecified. To guarantee a barotropic fluid that does not change the functional form of the Friedmann equations\footnote{Remarkably, we assume the energy-momentum tensor can be split into a linear sum of constituents. For Cardassian dark energy models \cite{cardassian, cardassian1} or semi-quantum scenarios \cite{semiquantum} this condition is violated. However, by virtue of the most recent experimental findings provided by the Planck satellite \cite{planck}, we presume the simplest hypothesis that appears statistically favored.}, the following requirements can be imposed:

\begin{equation}
\left\{
                 \begin{array}{ll}
                   G(z)\rightarrow 1 \,,\quad & \hbox{$z=0$;} \\
                   \,\\
                   \Omega_{DE}\equiv1-\Omega_{m}-\Omega_k\,,\quad & \hbox{$\forall z$;} \\
                   \,\\
                   \displaystyle\frac{G(z)}{(1+z)^3} \gtrsim\frac{\Omega_{m}}{\Omega_{DE}}\,,\quad & \hbox{$z\rightarrow 0$.}
                 \end{array}
               \right.
\end{equation}

Evidently, the first condition constraints $H=H_0$ at $z=0$, whereas the second  appears as a direct consequence of the first. The last implies that dark energy tends to dominate over matter at late-times.

Immediately, one can compute the cosmographic terms, $q$ and $j$ and obtain
\begin{widetext}
\begin{equation}\label{qudef}
q(z)=-1+\frac{(1+z)\left[3\Omega_{m}(1+z)^{2}+\Omega_{DE}G'(z)\right]}{2\left[\Omega_{m}(1+z)^{3}+\Omega_{DE}G(z)\right]}\,,
\end{equation}
and
\begin{equation}\label{jeidef}
j(z)=\frac{2\Omega_{DE}G(z)+(1+z)\left\{-2\Omega_{DE}G'(z)+(1+z)\left[2\Omega_{m}(1+z)+\Omega_{DE}G''(z)\right]\right\}}{2\left[\Omega_{m}(1+z)^{3}+\Omega_{DE}G(z)\right]}\,,
\end{equation}
\end{widetext}
where the superscript $'$ represents the derivative with respect to the redshift $z$ and so $G^\prime(z)=\frac{dG(z)}{dz}$.

The above quantities are general and can be calculated at any redshift. Our goal is to find a correspondence between $q_0$ and $j_0$, which represent the present values of the deceleration and jerk terms, respectively, and $q_{eq}$ and $j_{eq}$, which represent the values at the equivalence redshift $z_{eq}$. Our aim is to determine new constraints on $q_{eq}$ and $j_{eq}$ without assuming any specific form for the function $G(z)$.

For the sake of completeness, two specific times of interest in dark energy evolution are

\begin{itemize}
    \item[-] the transition redshift, which represents the onset of cosmic acceleration \cite{luongomuccinogrb};
    \item[-] the equivalence redshift, which can be obtained by invoking the equivalence between dark matter or baryons and dark energy.
\end{itemize}

The former has been extensively investigated in a model-independent way, see e.g. \cite{modelind, jesus}, while the latter has received relatively less attention in current literature. The challenge with the equivalence redshift is that both $q$ and $j$ do not vanish, as they do at the transition time, and the value of $q_{eq}$ is unknown, implying a further coefficient to be constrained.

Thus, to relate $q_0$ and $j_0$ to $q_{eq}$ and $j_{eq}$, we replace the mass density with the values of $q_0$ and $j_0$ using Eqs. (\ref{qudef}) and (\ref{jeidef}). We then substitute this value into the formal expression for $z_{eq}$ obtained in the context of a generic model, as shown in Eq. \eqref{hz}. This strategy removes one degree of freedom, \textit{i.e.}, taking the mass away from our fitting procedure and, furthermore, reduces the degeneracy problem cited above.

\section{Cosmographic reconstruction of the equivalence redshift}\label{sez3}

The equivalence redshift is frequently associated with the more popular equivalence between radiation and matter, that however occurs at the very early universe. At late-times,  there is the need to formulate the existence of a further redshift, roughly corresponding to the equivalence between dark energy and matter.
This occurs by virtue of the functional form of matter and dark energy. So, one would expect that such an  equivalence time can be used as \emph{cosmographic discriminator} toward the understanding of the right cosmological background, \textit{i.e.}, to clarify whether and how much dark energy evolves. At our time, then, the universe appears well-approximated by two fluids, matter and dark energy with the additional fluid-like contribution of spatial curvature, $\rho_k$.

Generally, invoking the total Hubble rate prompted in Eq. \eqref{hz}, the equivalence redshift fulfills the following recipe

\begin{equation}\label{formal}
\Omega_{DE}G(z_{eq})=\Omega_{m}(1+z_{eq})^3+\Omega_k(1+z_{eq})^2\,,
\end{equation}
and appears particularly interesting since
\begin{itemize}
    \item[-] it occurs \emph{at smaller redshifts} than transition between dark energy and dark matter, \textit{i.e.}, it is close to $z\simeq0$,
    \item[-] it has the advantage that\footnote{Contrary to the transition redshift, $z_{tr}$, \textit{i.e.}, the redshift at which the universe acceleration changes its sign, where $q(z_{tr})=0$ \cite{modelind}.} $q(z_{eq})\neq0$. Thus, just at low terms of the series in Eq. \eqref{serie1}, we can get constraints.
\end{itemize}
Below, we distinguish two cases. The first fixes the spatial curvature to be zero, namely assuming a spatially-flat universe, while the second fixes the value of curvature by using the measures made by the Planck satellite \cite{planck}. Consequently, we  compare later both the approaches to see whether the spatially-flat universe is statistically  favored in predicting the equivalence redshift than non-flat cosmologies.  For the sake of completeness, we remark that if one does not fix the spatial curvature today, the measurements of free parameters will be clearly degenerate with it. Nevertheless, a more complete approach, leaving $\Omega_k$ free to vary,  could be also investigated in future efforts.

\subsection{Spatially flat universe}

By solving  Eq. \eqref{formal}, in a spatially-flat universe, we have
\begin{eqnarray}\label{zeq1}
z_{eq}= \left[G(z_{eq})\frac{ \Omega_{DE}}{\Omega_{m}}\right]^{\frac{1}{3}}-1\,.
\end{eqnarray}

Immediately, one notices that the function $G(z_{eq})\gtrsim1$, since our starting hypothesis is to consider a barotropic fluid. Consequently, any complicate dark energy model would contribute by powers of $\sim (1+z)$, that is clearly larger than unity for any redshift. This condition reinforces $z_{eq}>0$, \textit{i.e.}, that the equivalence occurred in the far past. Clearly, the case $G(z_{eq})=1$ corresponding to the standard $\Lambda$CDM paradigm appears as a suitable limiting case, giving viable results since by definition $\frac{\Omega_{DE}}{\Omega_m}>1$. By extension, if a fuzzy dark energy model would be invoked, where the condition $G(z_{eq})>1$ is somehow violated, then we have to require that the degenerate product between this function and the ratio of the two magnitudes, $\Omega_{DE}$ and $\Omega_m$, permits to have $z_{eq}>0$. We limit ourselves to the simplest case above stated. As already reported, the limit to the concordance background model, namely the $\Lambda$CDM paradigm, gives:

\begin{equation}\label{zeqlcdm}
z_{eq,\Lambda}= \left(\frac{\Omega_\Lambda}{\Omega_{m}}\right)^{\frac{1}{3}}-1\,.
\end{equation}
Analogously, characterizing dark energy adopting the $\omega$CDM model and the Chevallier-Polarski-Linder (CPL)\footnote{To derive the expression for the equivalence redshift in the CPL parametrization, we perform a Taylor expansion up to the second order around $z_{eq}=0$. The CPL parametrization is given by $\omega=\omega_0+\omega_1\frac{z}{1+z}$. } parametrization \cite{CPL1, CPL2}, one gets respectively

\begin{align}\label{zeqwcdmCPL}
    &z_{eq,\omega}= \left(\frac{\Omega_{DE}}{\Omega_{m}}\right)^{-\frac{1}{3\omega}}-1,\\
    &z_{eq}^{CPL}\simeq \nonumber\\
    &\frac{3\omega_1\Omega_{DE}-3\Omega_{m}(\omega_0+\omega_1)-\sqrt{\left[3\Omega_{m}(\omega_0+\omega_1)-3\omega_1\Omega_{DE}\right]^2-2(\Omega_{DE}-\Omega_{m})\left[3\omega_1(3\omega_1+2)\Omega_{DE}-3(\omega_0+\omega_1)(3\omega_0+3\omega_1+1)\Omega_{m}\right]}}{3\omega_1(3\omega_1+2)\Omega_{DE}-3(\omega_0+\omega_1)(3\omega_0+3\omega_1+1)\Omega_{m}}\,.
\end{align}

Considering numerical constraints\footnote{For the concordance background and CPL, we employ the Planck results \cite{planck} for mass and $\omega$, say $\Omega_{m}=0.31\pm0.0073$ and $\omega=-0.96\pm 0.080$. For the CPL case we employ $\omega_0=-0.96\pm 0.080$ and $\omega_1=-0.29^{+0.32}_{-0.26}$. The $1\sigma$ error bars are computed adopting the standard logarithm rule of error propagation. The asymmetric error bars in Eq. \eqref{zeqmodels} for $z_{eq,CPL}$ are due to the errors employed on $\omega_1$.  }, we easily get

\begin{subequations}\label{zeqmodels}
    \begin{align}\label{zeqLCDM}
    z_{eq,\Lambda}&= 0.29\pm 0.015\,,\\
z_{eq,\omega}&= 0.31\pm 0.045\,,\\
z_{eq,CPL}&\simeq 0.31^{+0.046}_{-0.045}\,.
    \end{align}
\end{subequations}

At this stage, employing Eq. \eqref{qudef} and plugging $z=0$ gives us

\begin{equation}\label{qu0generico}
q_0=-1+\frac{3\Omega_{m}+\Omega_{DE}G^\prime_0}{2}\,,
\end{equation}
where $q_0\equiv q(z=0)$ and $\Omega_{DE}\equiv 1-\Omega_{m}$. Hence, inverting Eq. \eqref{zeq1} and plugging the corresponding mass density into Eq. \eqref{qu0generico}, we infer 

\begin{equation}\label{q0zeq}
q_0=\frac{1}{2}\frac{G(z_{eq}) -  (1 + z_{eq})^3(2G_0- G^\prime_0)}{G(z_{eq}) + G_0(1 + z_{eq})^3}\,,
\end{equation}

that depends upon two additional offsets, namely $G(z_{eq})$ and $G^\prime_0$, but \emph{it does not depend on the explicit value of the cosmic mass}, $\Omega_m$. As a matter of comparison, for the $\Lambda$CDM, $\omega$CDM and CPL cases, we simply obtain

\begin{subequations}\label{q0zeqmodels}
\begin{align}\label{q0zeqLCDM}
q_{0,\Lambda}&=\frac{1}{2}\frac{1 - 2 (1 + z_{eq})^3 }{1 + (1 + z_{eq})^3}\,.   \\
q_{0,\omega}&=\frac{1}{2}\frac{1+3\omega+(1+z_{eq})^{3\omega}}{1+(1+z_{eq})^{3\omega}}\,.\\
q_{0, CPL}&=\frac{1}{2}\left[\frac{(1+z_{eq})^{3(\omega_0+\omega_1)}\exp\left(-\frac{3\omega_1z_{eq}}{1+z_{eq}}\right)+3\omega_0+1}{(1+z_{eq})^{3(\omega_0+\omega_1)}\exp\left(-\frac{3\omega_1z_{eq}}{1+z_{eq}}\right)+1}\right]\,.
\end{align}
\end{subequations}

In the following graphic we plot Eqs. \eqref{q0zeqmodels} versus the equivalence redshift taking as constraints the values inside Eq. \eqref{zeqmodels} within the errors. 

\begin{figure}[H]
    \centering
    \includegraphics{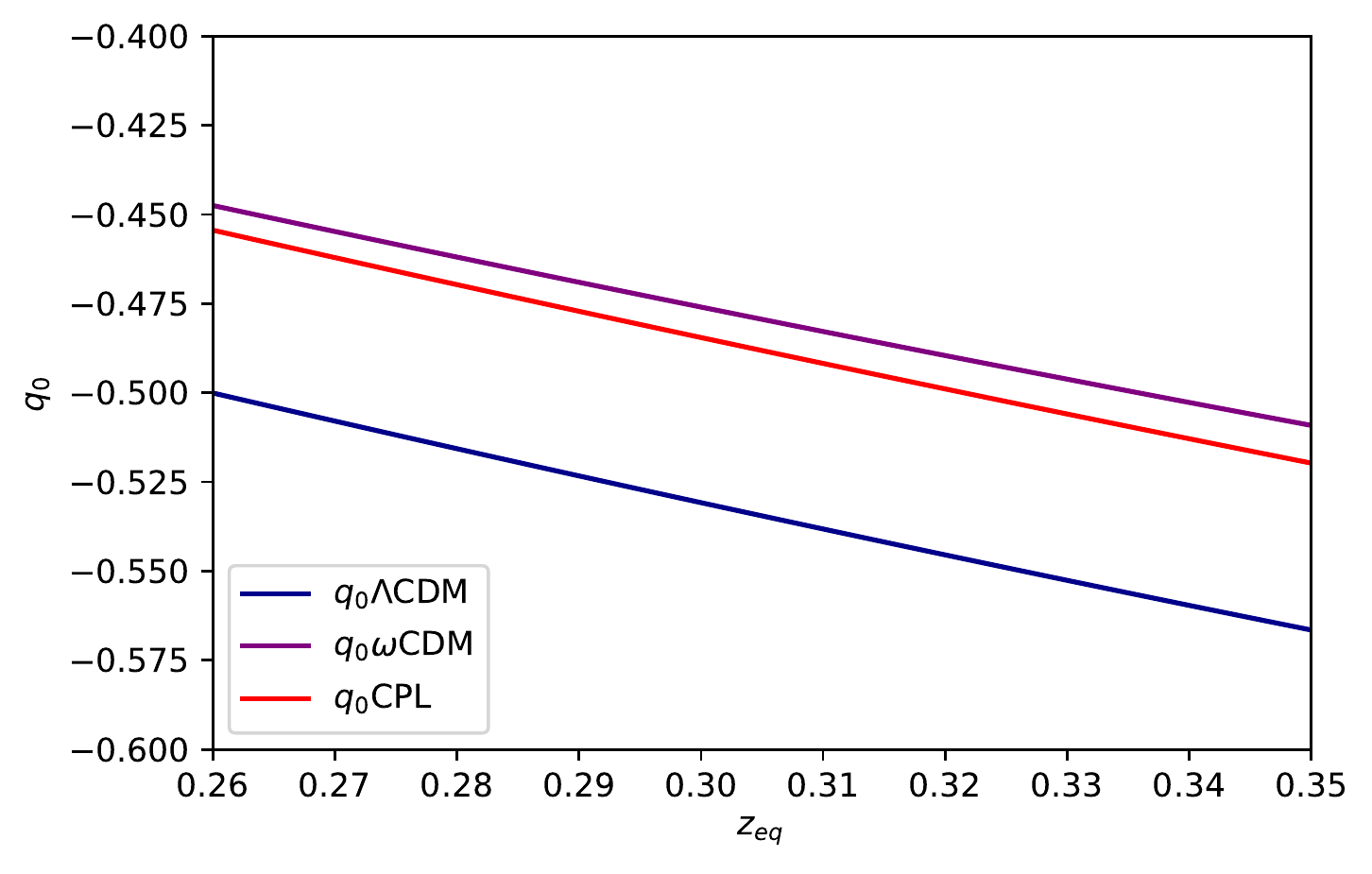}
    \caption{Plot of the deceleration parameter in the $\Lambda$CDM, $\omega$CDM model and CPL parametrization versus the equivalence redshift. For the equivalence redshift we consider $z_{eq}\in[0.26,0.35]$ and we take $\omega = -0.96$ for the $\omega$CDM model, $\omega_0 = -0.96$ and $\omega_1 = -0.29$ for the CPL parametrization, adopting the Planck constraints \cite{planck}. In order to obtain the interval of the equivalence redshift we calculate the sums and differences between the equivalence redshifts in Eqs. \eqref{zeqmodels} and their errors. Afterwards, we took the maximum and minimal value of the redshifts to be used as our intervals for $z_{eq}$.}
    \label{qvszeqflat}
\end{figure}

Following the same strategy, assuming Eq. \eqref{jeidef}, we infer

\begin{equation}\label{j0generico}
j_0=\frac{1}{2}
\frac{ 2(G_0 + \Omega_{m} (1 - G_0 + G^\prime_0) -
    G^\prime_0) + (1 - \Omega_{m}) (G^{\prime\prime}_0)}{ \Omega_{m} +  (1-\Omega_{m}) G_0}
\end{equation}
and consequently

\begin{equation}\label{j0zeq}
j_0=\frac{1}{2}\frac{2 G(z_{eq}) + (1 + z_{eq})^3 (2 G_0 -
    2 G^\prime_0 + G^{\prime\prime}_0)}{ G(z_{eq}) + (1 + z_{eq})^3 G_0}\,,
\end{equation}

\noindent and, again, as a matter of comparison, we obtain

\begin{subequations}\label{j0zeqmodels}
\begin{align}\label{q0zeqLCDM}
j_{0,\Lambda}&=1\,,\\
j_{0,\omega}&=\frac{1}{2}\frac{9\omega(1+\omega)+2(1+z_{eq})^{3\omega}+2}{1+(1+z_{eq})^{3\omega}}\,,\\
j_{0, CPL}&= \frac{1}{2}\left[\frac{2(1+z_{eq})^{3(\omega_0+\omega_1)}\exp\left(-\frac{3\omega_1z_{eq}}{1+z_{eq}}\right)+2+9\omega_0(1+\omega_0)+3\omega_1}{(1+z_{eq})^{3(\omega_0+\omega_1)}\exp\left(-\frac{3\omega_1z_{eq}}{1+z_{eq}}\right)+1}\right]\,.
\end{align}
\end{subequations}

Here, we plot Eqs. \eqref{j0zeqmodels} versus the equivalence redshift taking as constraints the values inside Eq. \eqref{zeqmodels} within the errors.

\begin{figure}[H]
    \centering
    \includegraphics{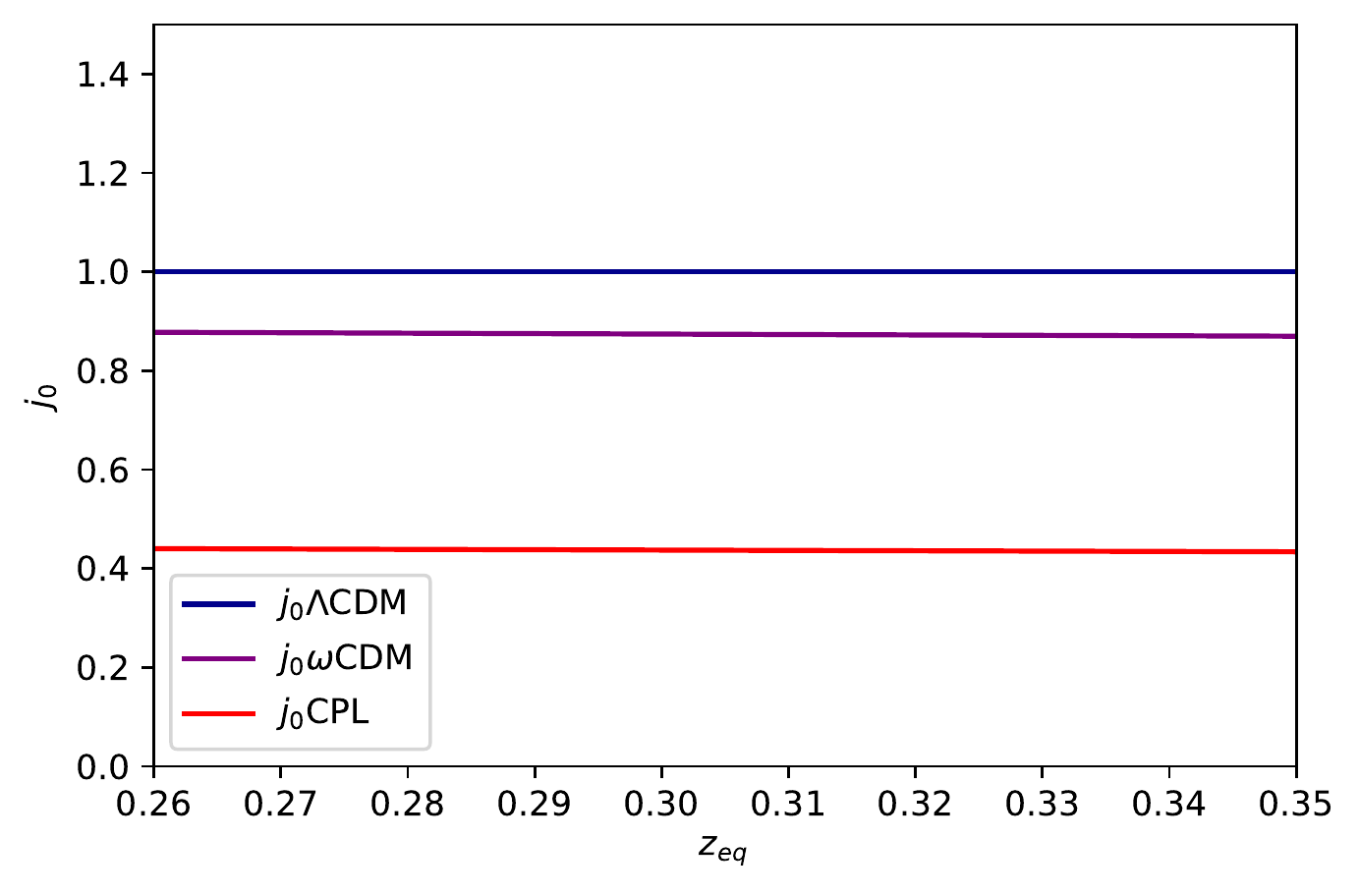}
    \caption{Plot of the jerk parameter in the $\Lambda$CDM, $\omega$CDM model and CPL parametrization versus the equivalence redshift. For the equivalence redshift we consider $z_{eq}\in [0.26,0.35]$ and we take $\omega = -0.96$ for the $\omega$CDM model, $\omega_0 = -0.96$ and $\omega_1 = -0.29$ for the CPL parametrization, adopting the Planck constraints \cite{planck}. In order to obtain the interval of the equivalence redshift we calculate the sums and differences between the equivalence redshifts in Eqs. \eqref{zeqmodels} and their errors. Afterwards, we took the maximum and minimal value of the redshifts to be used as our intervals for $z_{eq}$.}
    \label{jvszeqflat}
\end{figure}

\noindent In the above, we determined $q_0=q_0(G({z_{eq}}),G_0,G^\prime_0)$ to substitute into the expansions of cosmic distances and then to directly fit with data. In addition to the above, we analyze the effects of spatial curvature in the incoming subsection.

\subsection{Non-spatially flat universe}

Following the same strategy as for the spatially flat solution, we argue for a  non spatially-flat universe,

\begin{equation}\label{zeqnonflat}
z_{eq}=-1 + \frac{1}{6 \Omega_{m}} \left(  \frac{2^{\frac{4}{3}} \Omega_k^2}{y[\Omega_{m},\Omega_{k}]} + 2^{\frac{2}{3}}\cdot y[\Omega_{m},\Omega_{k}]-2 \Omega_k\right)\,,
\end{equation}
where the function $y[\Omega_{m},\Omega_{k}]$ is given by
\begin{align}
y[\Omega_{m},\Omega_{k}]&=\Big(-2 \Omega_k^3 - 27 G[z_{eq}] \Omega_m^2 (-1 + \Omega_k + \Omega_m) +\nonumber\\
&  3^{\frac{3}{2}} \sqrt{G[z_{eq}] \Omega_m^2 (-1 + \Omega_k + \Omega_m) (4 \Omega_k^3 +
        27 G[z_{eq}] (-1 + \Omega_k) \Omega_m^2 + 27 G[z_{eq}] \Omega_m^3)}\Big)^\frac{1}{3}\,.
\end{align}

We can derive a form for the equivalence redshift in the non-flat scenario considering the cosmological paradigm, $\omega$CDM model and CPL parametrization. In the case of the $\Lambda$CDM model we have

\begin{eqnarray}
    z_{eq,\Lambda} = -1+\frac{1}{6\Omega_m}\left(\frac{2^{\frac{4}{3}}\Omega_k^2}{y[\Omega_m, \Omega_k, \Omega_\Lambda]}+2^{\frac{2}{3}}y[\Omega_m, \Omega_k, \Omega_\Lambda]-2\Omega_k\right),
\end{eqnarray}

where

\begin{align}
        y(\Omega_m, \Omega_k, \Omega_\Lambda)&= \left[-2\Omega_k^3+27\Omega_m^2\Omega_\Lambda+3^{\frac{3}{2}}\sqrt{\Omega_m^2\Omega_\Lambda(-4\Omega_k^3+27\Omega_m^2\Omega_\Lambda)}\right]^\frac{1}{3}.
    \end{align}

In the case of the $\omega$CDM model we get\footnote{After expanding $\Omega_{DE}(1+z_{eq})^{3(\omega+1)}=\Omega_m(1+z_{eq})^3+\Omega_k(1+z_{eq})^2$ in a Taylor series around $z_{eq}=0$ up to the first order.}

\begin{eqnarray}
    z_{eq,\omega} \simeq \frac{\Omega_k+\Omega_m-\Omega_{DE}}{3\omega\Omega_{DE}-\Omega_k}.
\end{eqnarray}

For the CPL parametrization we have\footnote{After expanding $\Omega_{DE}(1+z_{eq})^{3(\omega_0+\omega_1+1)}\exp\left(-\frac{3\omega_1z_{eq}}{1+z_{eq}}\right)=\Omega_m(1+z_{eq})^3+\Omega_k(1+z_{eq})^2$ in a Taylor series around $z_{eq}=0$ up to the second order.}

\begin{eqnarray}
    z_{eq,CPL}\simeq \frac{3\omega_1\Omega_{DE}-3(\omega_0+\omega_1)(\Omega_m+\Omega_k)-\Omega_k-x(\Omega_m, \Omega_{DE}, \Omega_k, \omega_0, \omega_1)}{3\omega_1\Omega_{DE}(3\omega_1+2)-(3\omega_0+3\omega_1+1)[3(\omega_0+\omega_1)(\Omega_m+\Omega_k)+2\Omega_k]},
\end{eqnarray}

where

\begin{eqnarray}
     x(\Omega_m, \Omega_{DE}, \Omega_k, \omega_0, \omega_1) = \{[3(\omega_0+\omega_1)(\Omega_m+\Omega_k)+\Omega_k-3\omega_1\Omega_{DE}]^2-2(\Omega_{DE}-\Omega_m-\Omega_k)[3\omega_1(3\omega_1+2)\Omega_{DE}+\\-(3\omega_0+3\omega_1+1)(3\Omega_m(\omega_0+\omega_1)+\Omega_k(3\omega_0+3\omega_1+2)]\}^\frac{1}{2}. \nonumber
\end{eqnarray}

In analogy to the procedure developed for the flat case, considering numerical constraints, we get

\begin{subequations}\label{zeqmodelsnonflatnum}
    \begin{align}\label{zeqLCDMnonflat}
    z_{eq,\Lambda}&= 0.36^{+0.048}_{-0.037},\\
    z_{eq,\omega}&\simeq 0.22^{+0.042}_{-0.036}\,,\\
    z_{eq,CPL}&\simeq 0.38^{+0.095}_{-0.080}.
    \end{align}
\end{subequations}

The deceleration and jerk parameter in a non-flat scenario can be reobtained from Eq. \eqref{hz} as

\begin{equation}\label{qnonflat}
    q(z) = -1+\frac{(1+z)[3\Omega_m(1+z)^2+2\Omega_k(1+z)+\Omega_{DE}G'(z)]}{2[\Omega_m(1+z)^3+\Omega_k(1+z)^2+\Omega_{DE}G(z)]},
\end{equation}

and

\begin{equation}\label{jnonflat}
    j(z) = 1+\frac{(1+z)[\Omega_{DE}G''(z)(1+z)-2\Omega_{DE}G'(z)-2\Omega_k(1+z)]}{2[\Omega_m(1+z)^3+\Omega_k(1+z)^2+\Omega_{DE}G(z)]}.
\end{equation}

Spatial curvature is approximately compatible with zero. Current Planck value furnishes \cite{planck}

\begin{equation}\label{spatcurv}
\Omega_k^{Pl}=-0.056^{+0.028}_{-0.018}\,,
\end{equation}
indicating that, in order to exploit the curvature term from the previous expressions, one can expand up to the first order Eqs. \eqref{qnonflat}-\eqref{jnonflat} around $\Omega_k=0$. Thus, we approximate Eqs. \eqref{qnonflat}-\eqref{jnonflat} by

\begin{eqnarray}\label{qtilde}
    {q(z)} = \ -1&+&\frac{(1+z)[3\Omega_m(1+z)^2+(1-\Omega_m)G'(z)]}{2[\Omega_m(1+z)^3+(1-\Omega_m)G(z)]}+\\&+&\frac{(1+z)^2\Omega_k\left(-\Omega_m(1+z)^3+G(z)[2+\Omega_m(3z+1)]-(1+z)(\Omega_mz+1)G'(z)\right)}{2[\Omega_m(1+z)^3+(1-\Omega_m)G(z)]^2},\nonumber
\end{eqnarray}

and

\begin{eqnarray}\label{jtilde}
    {j(z)} = \nonumber 1&+&\frac{(1+z)[(1-\Omega_m)G''(z)(1+z)-2(1-\Omega_m)G'(z)]}{2[\Omega_m(1+z)^3+(1-\Omega_m)G(z)]}+\\&+&\frac{\Omega_k(1+z)^2}{2[\Omega_m(1+z)^3+(1-\Omega_m)G(z)]^2}[-\Omega_mG''(z)(1+z)^3+2\Omega_mG'(z)(1+z)^2-2\Omega_m(1+z)^3+\\&-&2(1-\Omega_m)G(z)-(1-\Omega_m)G''(z)(1+z)^2+2(1-\Omega_m)G'(z)(1+z)],\nonumber
\end{eqnarray}

and therefore finding $\Omega_m$ out from Eqs. \eqref{qtilde}-\eqref{jtilde}, having imposed $z=0$, we substitute the mass value inside Eq. \eqref{zeqnonflat}. This strategy, analogous to the previous one, however, does not allow us to perform exact computation. So, together with the above Taylor expansions, we set $\Omega_k\approx -\frac{\Omega_m}{6}$ in order to simplify\footnote{Additional clarifications on how the calculations have been carried out to simplify $z_{eq}$ in the non-flat scenario are given in Appendix \ref{zeqOkdiv0}.} Eq. \eqref{zeqnonflat} leading to

\begin{equation}\label{qtildenonflat}
    {q_0} \simeq \frac{(1+z_{eq})^3(G^\prime_0-2)(1-\Omega_k)-G(z_{eq})(G^\prime_0-3)}{2(1+z_{eq})^3}.
\end{equation}

As a matter of comparison we consider the $\Lambda$CDM, $\omega$CDM and CPL case obtaining

\begin{subequations}\label{qnonflat3}
    \begin{align}
    q_{0,\Lambda}&\simeq \frac{3-2(1+z_{eq})^3(1-\Omega_k)}{2(1+z_{eq})^3} \,,\\
     q_{0, \omega}&\simeq \frac{(1-\Omega_k)(3\omega+1)-3\omega(1+z_{eq})^{3\omega}}{2}\,,\\
    q_{0, CPL}&\simeq \frac{(1-\Omega_k)(3\omega_0+1)-3\omega_0(1+z_{eq})^{3(\omega_0+\omega_1)}\exp\left(-\frac{3\omega_1z_{eq}}{1+z_{eq}}\right)}{2}.
    \end{align}
\end{subequations}

Here, we plot Eqs. \eqref{qnonflat3} versus the equivalence redshift taking as constraints the values inside Eq. \eqref{zeqmodelsnonflatnum} within the errors. 

\begin{figure}[H]
    \centering
    \includegraphics{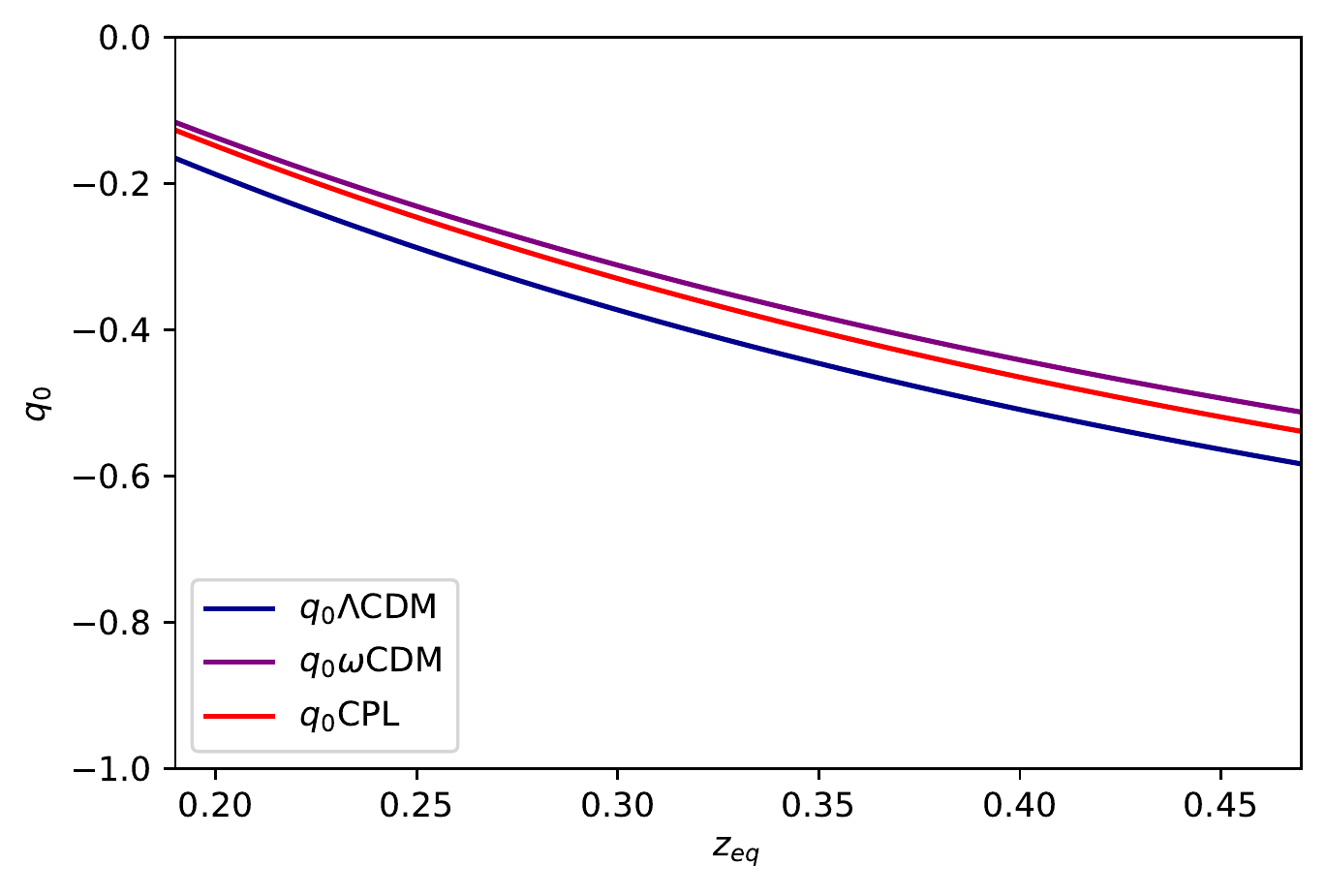}
    \caption{Plot of the deceleration parameter in the $\Lambda$CDM, $\omega$CDM model and CPL parametrization versus the equivalence redshift in the non-flat scenario. For the equivalence redshift we consider $z_{eq}\in [0.19,0.47]$ and we take $\omega = -0.96$ for the $\omega$CDM model, $\omega_0 = -0.96$ and $\omega_1 = -0.29$ for the CPL parametrization, adopting the Planck constraints \cite{planck}. In order to obtain the interval of the equivalence redshift we calculate the sums and differences between the equivalence redshifts in Eqs. \eqref{zeqmodelsnonflatnum} and their errors. Afterwards, we took the maximum and minimal value of the redshifts to be used as our intervals for $z_{eq}$.}
    \label{qvszeqnonflat}
\end{figure}

Following the same approximation we did for Eq. \eqref{qtildenonflat} for the jerk parameter we obtain

\begin{equation}\label{jtildenonflat}
    {j_0} \simeq \frac{(1+z_{eq})^3[2(1-G^\prime_0)+G^{\prime\prime}_0]-\Omega_k(1+z_{eq})^3[2(1-G^\prime_0)-G^{\prime\prime}_0]-G(z_{eq})[G^{\prime\prime}_0-2G^\prime_0]}{2(1+z_{eq})^3}.
\end{equation}

Again, we get for our chosen models

\begin{subequations}\label{jnonflat3}
\begin{align}
    j_{0,\Lambda}&= 1-\Omega_k \,,\\
    j_{0, \omega}&\simeq \frac{2+9\omega(\omega+1)+\Omega_k[10+3\omega(3\omega+7)]-9\omega(\omega+1)(1+z_{eq})^{3\omega}}{2}\,,\\
    j_{0,CPL}&\simeq \frac{1}{2}\Bigg(2+9\omega_0(\omega_0+1)+3\omega_1+\Omega_k[10+3\omega_0(3\omega_0+7)+3\omega_1]+\\&-(1+z_{eq})^{3(\omega_0+\omega_1)}\exp\left(-\frac{3\omega_1z_{eq}} {1+z_{eq}}\right)[9\omega_0(\omega_0+1)+3\omega_1]\Bigg) \nonumber.
\end{align}

\end{subequations}

Here, we plot Eqs. \eqref{jnonflat3} versus the equivalence redshift taking as constraints the values inside Eq. \eqref{zeqmodelsnonflatnum} within the errors. 

\begin{figure}[H]
    \centering
    \includegraphics{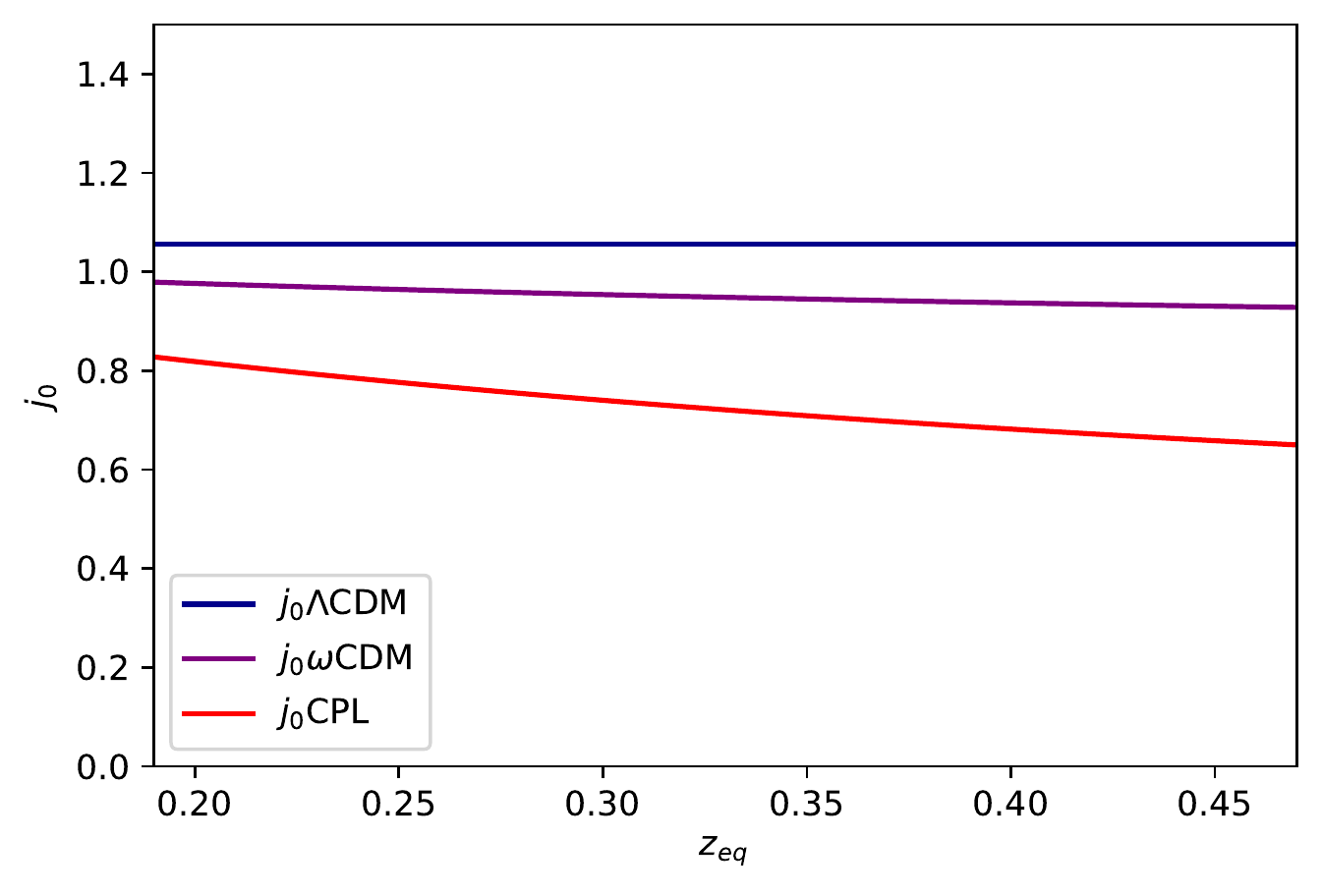}
    \caption{Plot of the jerk parameter in the $\Lambda$CDM, $\omega$CDM model and CPL parametrization versus the equivalence redshift in the non-flat scenario. For the equivalence redshift we consider $z_{eq}\in [0.19,0.47]$ and we take $\omega = -0.96$ for the $\omega$CDM model, $\omega_0 = -0.96$ and $\omega_1 = -0.29$ for the CPL parametrization, adopting the Planck constraints \cite{planck}. In order to obtain the interval of the equivalence redshift we calculate the sums and differences between the equivalence redshifts in Eqs. \eqref{zeqmodelsnonflatnum} and their errors. Afterwards, we took the maximum and minimal value of the redshifts to be used as our intervals for $z_{eq}$.}
    \label{jvszeqnonflat}
\end{figure}

Furthermore, we plot in Figs. \ref{qvszeqflat}, \ref{jvszeqflat},  \ref{qvszeqnonflat} and \ref{jvszeqnonflat} the cosmographic parameters, namely the deceleration and jerk parameter versus the equivalence redshift for three cosmological models: the $\Lambda$CDM, $\omega$CDM and CPL parametrization in both the cases of flat and non-flat universe.

Specifically, in Figs. \ref{qvszeqflat}, \ref{jvszeqflat}, we observe that the deceleration parameter tends to small values as the equivalence redshift increases. Nevertheless, in the same case, the jerk parameter appears approximately stable, with the clear exception of the $\Lambda$CDM model, where its value is exactly $j=1$ at all time. 

This appears quite different in the non-flat case. Indeed, focusing on Figs. \ref{qvszeqnonflat} and \ref{jvszeqnonflat}, we observe a similar trend for the deceleration parameter, slightly different due to the presence of spatial curvature, that, instead, is more evident for the jerk parameter. There, again with the exception of the $\Lambda$CDM case, where the corresponding value is $j_0=1-\Omega_k$, the other quantities tend to decrease more deeply than the flat case, as we expected.

\subsection{Other equivalence redshift definitions}

The definition of the equivalence redshift assumes that dark energy and matter equate at a given redshift. Following the same procedure developed to define $z_{eq}$ above, we can therefore notice that two more equivalence times can be associated with

\begin{align}\label{formal2}
\Omega_{DE}G(z_{eq})&=\Omega_{cdm}(1+z_{eq})^3+\Omega_k(1+z_{eq})^2\,,\\
\Omega_{DE}G(z_{eq})&=\Omega_{b}(1+z_{eq})^3+\Omega_k(1+z_{eq})^2\,, \label{formal3}
\end{align}

where $\Omega_m\equiv\Omega_{cdm}+\Omega_b$, with $\Omega_{cdm}$ the cold dark matter density and $\Omega_b$ the baryonic density. Thus, Eqs. \eqref{formal2}-\eqref{formal3} in the flat case lead to

\begin{subequations}\label{zeq1ALTRO}
    \begin{align}
     z_{eq,cdm}&= \left[G(z_{eq})\frac{ \Omega_{DE}}{\Omega_{cdm}}\right]^{\frac{1}{3}}-1\,,\\
z_{eq,b}&= \left[G(z_{eq})\frac{ \Omega_{DE}}{\Omega_{b}}\right]^{\frac{1}{3}}-1\,.
    \end{align}
\end{subequations}

Hence, a direct combination between Eqs. \eqref{zeq1} and \eqref{zeq1ALTRO} provides

\begin{subequations}\label{combinatio}
    \begin{align}
     z_{eq,cdm}&= -1+\left(\frac{\Omega_m}{\Omega_{cdm}}\right)^{\frac{1}{3}}(1+z_{eq})\,,\\
z_{eq,b}&= -1+\left(\frac{\Omega_m}{\Omega_b}\right)^{\frac{1}{3}}(1+z_{eq})\,.
    \end{align}
\end{subequations}

Considering numerical constraints\footnote{For the matter density we employ as usual the one from Planck. Regarding the cold dark matter and baryon density we consider that $\Omega_{cdm} \sim \frac{5}{6}\Omega_m = 0.26\pm 0.0061$ while $\Omega_b \sim \frac{1}{6}\Omega_m = 0.052\pm 0.0012$. For the equivalence redshift we employ the one from Eq. \eqref{zeqLCDM}. In the non-flat case we employ the same constraints together with the spatial curvature from Eq. \eqref{spatcurv}.}, we get

\begin{subequations}
    \begin{align}
       z_{eq,cdm}&= 0.38\pm 0.038\,,\\
       z_{eq,b}&= 1.35\pm 0.063\,.
    \end{align}
\end{subequations}

In analogy, the same equivalence redshifts of baryons and cold dark matter, $z_{eq,b}$ and $z_{eq,cdm}$ occur as spatial curvature is included\footnote{For more clarifications on how the computations have been carried out, we refer to Appendix \ref{zeqOkdiv01}.}, giving

\begin{subequations}\label{combinatio2}
    \begin{align}
     z_{eq,cdm}&= -1-\frac{\Omega_k}{3\Omega_{cdm}}+\frac{1}{3\Omega_{cdm}2^{\frac{1}{3}}}\left[\left(a(\Omega_k, \Omega_m, \Omega_{cdm})+b(\Omega_k, \Omega_m, \Omega_{cdm})\right)^{\frac{1}{3}}+\left(a(\Omega_k, \Omega_m, \Omega_{cdm})-b(\Omega_k, \Omega_m, \Omega_{cdm})\right)^{\frac{1}{3}}\right]\,,\\
z_{eq,b}&= -1-\frac{\Omega_k}{3\Omega_b}+\frac{1}{3\Omega_b2^{\frac{1}{3}}}\left[\left(c(\Omega_k, \Omega_m, \Omega_b)+d(\Omega_k, \Omega_m, \Omega_b)\right)^{\frac{1}{3}}+\left(c(\Omega_k, \Omega_m, \Omega_b)-d(\Omega_k, \Omega_m, \Omega_b)\right)^{\frac{1}{3}}\right]\,,
    \end{align}
\end{subequations}

where

\begin{subequations}
    \begin{align}
        a(\Omega_k, \Omega_m, \Omega_{cdm}) &= -2\Omega_k^3+27\Omega_{cdm}^2(1+z_{eq})^2(\Omega_k+\Omega_m(1+z_{eq})),\\
        b(\Omega_k, \Omega_m, \Omega_{cdm}) &= 3^{\frac{3}{2}}\sqrt{\Omega_{cdm}^2(1+z_{eq})^2(\Omega_k+\Omega_m(1+z_{eq}))(-4\Omega_k^3+27\Omega_{cdm}^2(1+z_{eq})^2(\Omega_k+\Omega_m(1+z_{eq})))}, \\
        c(\Omega_k, \Omega_m, \Omega_b) &= -2\Omega_k^3+27\Omega_b^2(1+z_{eq})^2(\Omega_k+\Omega_m(1+z_{eq})),\\
        d(\Omega_k, \Omega_m, \Omega_b) &= 3^{\frac{3}{2}}\sqrt{\Omega_b^2(1+z_{eq})^2(\Omega_k+\Omega_m(1+z_{eq}))(-4\Omega_k^3+27\Omega_b^2(1+z_{eq})^2(\Omega_k+\Omega_m(1+z_{eq})))}.
    \end{align}
\end{subequations}

In the non-flat case considering Eq. \eqref{zeqLCDMnonflat} we achieve

\begin{subequations}
    \begin{align}
        z_{eq,cdm}&= 0.45^{+0.081}_{-0.067}\,,\\
        z_{eq,b}&=  1.77^{+0.32}_{-0.23}\,.\\ \nonumber
    \end{align}
\end{subequations}

\section{Data comparison from  cosmographic  expansions}\label{sec3}

To get limits over $q_0$ and $j_0$ in a model-independent way, one can simply consider the standard luminosity distance as function of the redshift as:

\begin{equation}\label{stp}
 d_L  = \frac{(1+z)H_0^{-1}}{\sqrt{|\Omega_k|}}S_k\left[\sqrt{|\Omega_k|}\int^{z}_0 \frac{H_0}{H(z')} dz'\right]\,,
\end{equation}
where

\begin{equation}
    S_k(x)= \begin{cases}
    \sin(x) & k=+1, \\
    x & k=0, \\
    \sinh(x) & k=-1.
  \end{cases}
\end{equation}

Rewriting the luminosity distance by means of Eq. \eqref{serie1} up to the third order cosmographic parameters furnishes\footnote{See Ref. \cite{luongoprd2012}, for more general cosmic expansions.}

\begin{equation}\label{lumdis1}
    d_{L}^{(3)} \approx   \frac{z}{H_0} \left( 1 + \alpha_1\,z + \alpha_2\,z^2 + \ldots\right)\,,
\end{equation}
where, the coefficients $\alpha_1$ and $\alpha_2$ depend on the spatial curvature, besides being function of the cosmographic series. For a spatially-flat universe, we get

\begin{subequations}
\begin{align}
\alpha_1(k=0)&=\frac{1}{2} - \frac{q_0}{2}\,,\label{lumdis2}\\
\alpha_2(k=0)&=\frac{q_0^2}{2}-\frac{1}{6}  + \frac{q_0}{6} -\frac{j_0}{6}\,.\label{lumdis3}
\end{align}
\end{subequations}

For a non-flat universe, instead we obtain

\begin{subequations}\label{dlnonflat1}
\begin{align}
\alpha_1(k\neq0)&=\alpha_1(k=0)\,,\label{lumdis2k}\\
\alpha_2(k\neq0)&=\alpha_2(k=0)-\frac{k}{6\rho_{c}a_0^2}\,,\label{lumdis3k}
\end{align}
\end{subequations}

where $\Omega_k\equiv-\frac{k}{H_0^2a_0^2}\frac{8\pi G}{3} $ and $\rho_{c}\equiv\frac{3H_0^2}{8\pi G}$.

It is evident that a direct comparison of $d_{L}^{(3)}$ with cosmological data permits to fix $q_0$ and $j_0$ in a model independent way. In fact, in Eq. (\ref{stp}), no information on the form of $H(z)$ have been introduced before expanding to get Eq. (\ref{lumdis1}). Clearly, observational measurements of $j_0$ and $q_0$ can be replaced by measuring $z_{eq}$ and the evolution of dark energy directly.

Our strategy consists of the following points.

\begin{itemize}
    \item[-] We substitute into $\alpha_1$ and $\alpha_2$ the quantities reported in Eqs. \eqref{j0generico} and \eqref{j0zeq}.
    \item[-] We compute the observable quantities built up in terms of the luminosity distance.
    \item[-] We constrain \emph{directly} with data, in a model independent way, binning the functions $G$ and its derivatives.
\end{itemize}

To get constraints on the equivalence redshift and the other observable quantities, \textit{i.e.}, $G(z_{eq})$, $G^\prime_0$ and $G^{\prime\prime}_0$ our strategy is to work out a Monte Carlo Markov chain (MCMC) adopting the Metropolis-Hastings algorithm \cite{metropol}\footnote{For a brief discussion on the concept of Markov chains along with
their central role to Monte Carlo simulations, we refer to Ref. \cite%
{binder2000}. For an explicit description of the problems to be solved with
Monte Carlo methods, we suggest Ref. \cite{mackay2003}. Finally, for a nice
conceptual introduction to MCMC methods \cite{hastings70} along with simple
illustrative examples, we refer to Ref. \cite{speagle20}.} within a Python code, modifying the free-available code from  Ref. \cite{bocquet} for plotting our contours.

In so doing we adopt three samples of data, involving different kinds of measurements reported in terms of the redshift $z$. The measurements are:
\begin{enumerate}
    \item 32 observational Hubble data (OHD)\footnote{The name of the catalog was first given by Refs. \cite{OHD1, OHD2, OHD3}} from Ref. \cite{kumar, zhang}, where we measure directly the Hubble rate value at different times;
    \item the 1048 Type Ia Supernovae (SNe Ia) from the Pantheon catalog \cite{scolnic} for the non-flat scenario and under the form of six $H_0/H(z)$ data points in Ref. \cite{riess} for the flat scenario\footnote{Here, for the flat case we adopt the  reduced catalog that has been demonstrated to decrease the complexity of computations. The catalog is fully-equivalent to the entire one, so provides the advantage to speed up the overall evaluations of our parameters. On the other hand, the same catalog cannot be used for the non-flat case.}. In both cases, we consider magnitudes versus redshifts;
    \item the eight uncorrelated baryonic acoustic oscillations (BAO) data points from Ref. \cite{luongomuccino}, considering the angular distance versus redshift measurements.
\end{enumerate}

For each of these data set we have a corresponding log-likelihood entering our computation. As it is well-known, since the measurements are uncorrelated, the total likelihood function is the sum of all the likelihoods. Hence, we below describe in detail the corresponding log-likelihood definition for each of the above catalogs.

\begin{itemize}
    \item[-] \textbf{OHD}: The 32 OHD measurements are determined through spectroscopic observations. This ensures that the measurements are model-independent and it is possible to directly get $H(z)$ at different $z$. To this end, the values of the Hubble parameter are obtained through cosmic chronometers, \textit{i.e.}, using the difference between the age of passively evolving red galaxies \cite{jimenez, moresco}.

    This recipe permits one to define
    \begin{eqnarray}
       H(z_n) = -\frac{1}{1+z_n}\frac{dz_n}{dt},
    \end{eqnarray}

where $z_n$ involves the redshift interval which separates two nearby galaxies formed at the same time.

Looking at the data set here involved, the redshift lies on the interval $[0.070; 1.965]$ and the OHD log-likelihood is
    \begin{equation}
        \ln\mathcal{L}_{OHD} = -\frac{1}{2}\sum^{N_{OHD}}_{n=1}\left\{\left[\frac{H(z_n)-H_{th}(z_n)}{\sigma_{H(z_n)}}\right]^2+\ln\left(2\pi\sigma_{H(z_n)}^2\right)\right\},
    \end{equation}
    where $N_{OHD}=32$ and $H(z_n)$ are the 32 OHD, $\sigma_{H(z_n)}$ their errors and $H_{th}(z_n)$ the theoretical Hubble rates computed at $z_n$.

    \item[-] \textbf{SNe Ia}: The Pantheon catalog we employ here is considered the largest samples of SNe Ia, combining 1048 sources \cite{scolnic}. In our computation we use all Pantheon sample for the non-flat case while for the flat scenario we employ the catalog in the reduced form of six data points \cite{riess}. Looking at the catalog of data, in the non-flat case the redshift spans within the interval $[0.0089;2.2596]$ while in the flat scenario the redshift is within the interval $[0.07;1.50]$. The SNe Ia log-likelihood in the flat scenario is
    \begin{equation}
        \ln\mathcal{L}_{SN} = -\frac{1}{2}\sum^{N_{SN}}_{n=1}\left[\frac{H_0}{H(z_n)}-\frac{H_0}{H_{th}(z_n)}\right]^{T}\textbf{C}_{SN}^{-1}\left[\frac{H_0}{H(z_n)}-\frac{H_0}{H_{th}(z_n)}\right]-\frac{1}{2}\sum^{N_{SN}}_{n=1}\ln\left(2\pi|\det\textbf{C}_{SN}|\right),
    \end{equation}
where $N_{SN}=6$ and $\textbf{C}_{SN}$ is the covariance matrix.

In order to get the log-likelihood in the non-flat case we first define the distance modulus \cite{scolnic}
    \begin{eqnarray}
        \mu = m_B-\mathcal{M}+\alpha\chi_1-\beta\mathcal{C}+\Delta_M+\Delta_B,
    \end{eqnarray}
where $m_B$ is the $B$-band apparent magnitude, $\alpha$ and $\beta$ are coefficients, the first is related to the luminosity-stretch relation $\chi_1$ while the second is related to the luminosity-color relation $\mathcal{C}$, $\mathcal{M}$ is the $B$-band absolute magnitude and $\Delta_M$ and $\Delta_B$ are distance corrections.

Considering that the uncertainties do not depend on $\mathcal{M}$, the $B$-band absolute magnitude is removed by marginalizing over it so that we have  \cite{conley}

    \begin{eqnarray}
        \ln\mathcal{L}_{SN} = -\frac{1}{2}\left(a+\log\frac{e}{2\pi}-\frac{b^2}{e}\right),
    \end{eqnarray}

    where $a=\Delta\Vec{\mu}_{P}^T\textbf{C}_P^{-1}\Delta\Vec{\mu}_{P}$, $b=\Delta\Vec{\mu}_{P}^T\textbf{C}_P^{-1}\textbf{1}$ and $c=\textbf{1}^T\textbf{C}_P^{-1}\textbf{1}$.

    In $a$ and $b$ $\Delta\mu_{P}=\mu_P-\mu_P^{th}(\textbf{x}, z)$ is the difference between the observed distance moduli from the Pantheon sample and the theoretical distance moduli. On the other hand $\textbf{C}_P^{-1}$ is the inverse of the covariance matrix from the Pantheon sample.

    \item[-] \textbf{BAO}: these measures are imprinted in the power spectrum of large-scale structures \cite{seoeisenstein}. These oscillations arise due to sound waves that cause baryonic material in a spherical shell of radius $r_s$, \textit{i.e.}, the comoving sound horizon to expand \cite{percival}. The BAO log-likelihood is

    \begin{equation}
        \ln\mathcal{L}_{BAO} = -\frac{1}{2}\sum^{N_{BAO}}_{n=1}\left\{\left[\frac{d_A(z_n)-d_A^{th}(z_n)}{\sigma_{d_A}}\right]^2+\ln\left(2\pi\sigma_{d_A}^2\right)\right\},
    \end{equation}

    where $N_{BAO}=8$ and $d_A(z_n)$ are the angular distances from Ref. \cite{riess} while $d_A^{th}(z_n)$ are the theoretical angular distances

    \begin{equation}\label{datheo}
    d_A^{th}(z) = r_s\left[\frac{H_{th}(z)}{z}\right]^{\frac{1}{3}}\left[\frac{(1+z)}{d_L(z)}\right]^{\frac{2}{3}},
    \end{equation}

    where $r_s=147.21\pm 0.48$ is the comoving sound horizon from Ref. \cite{planck}. We substitute inside the luminosity distance $d_L(z)$ Eq. \eqref{lumdis1}.

\end{itemize}

In all the log-likelihoods, $H_{th}$ is the theoretical Hubble rate expanded in terms of the cosmographic parameters up to the second order

\begin{equation}\label{hth}
    H_{th}^{(2)}(q_0, j_0, z) \simeq H_0\left[1+(1+q_0)z+\frac{z^2}{2}(j_0-q_0^2)\right],
\end{equation}

where we substitute $q_0$ and $j_0$ with Eqs. \eqref{q0zeq}-\eqref{j0zeq}, respectively.

In the following table we illustrate our results using the OHD+SN and OHD+SN+BAO data sets

\begin{table}[H]
    \centering
    \setlength{\tabcolsep}{1.em}
    \renewcommand{\arraystretch}{1.3}
    \begin{tabular}{cccccc} \hline\hline
    {Data set} & {$h_0$} & {$G(z_{eq})$} & {$G^\prime_0$} & {$G^{\prime\prime}_0$} & {$z_{eq}$} \\ \hline\hline
    {OHD+SN} & {$0.69^{+0.020(0.04)}_{-0.020(0.04)}$}  & {$1.09^{+2.02(2.81)}_{-0.77(0.90)}$} & {$0.31^{+0.47(0.66)}_{-1.44(1.44)}$} & {$0.44^{+0.56(0.56)}_{-0.36(0.36)}$} & {$0.56^{+0.14(0.14)}_{-0.36(0.36)}$} \\ \hline
    {OHD+SN+BAO} & {$0.69^{+0.01(0.02)}_{-0.01(0.02)}$}  &{$0.87^{+1.15(1.23)}_{-0.89(0.96)}$} & {$0.49^{+0.67(0.79)}_{-0.44(0.68)}$} & {$-0.36^{+1.36(1.36)}_{-0.64(0.64)}$} & {$0.46^{+0.24(0.24)}_{-0.26(0.26)}$} \\  \hline\hline

    \end{tabular}
    \caption{Best-fit parameters of the Hubble rate, the function $G$ and its derivatives at $z=0$ and the equivalence redshift at 1$\sigma$-2$\sigma$ for the OHD+SN and OHD+SN+BAO data sets in the case of a flat universe.}
    \label{bestfitflat}
\end{table}

In the context of a non-flat universe we use the same log-likelihoods as before with the exception that the luminosity distance up to the third order inside Eq. \eqref{datheo} is built up by means of Eqs. \eqref{lumdis1}-\eqref{dlnonflat1}

Another difference with the flat case is that inside Eq. \eqref{hth} we substitute the cosmographic parameters in the non-flat scenario, Eqs. \eqref{qtildenonflat}-\eqref{jtildenonflat}.

Thus, after performing a MCMC in the non-flat case, the following results using OHD+SN and OHD+SN+BAO data set are shown in the following table

\begin{table}[H]
    \centering
     \setlength{\tabcolsep}{1.em}
    \renewcommand{\arraystretch}{1.3}
    \begin{tabular}{cccccc} \hline\hline
    {Data set} & {$h_0$} & {$G(z_{eq})$} & {$G^\prime_0$} & {$G^{\prime\prime}_0$} & {$z_{eq}$} \\ \hline\hline
    {OHD+SN} & {$0.71^{+0.017(0.036)}_{-0.018(0.035)}$}  & {$1.08^{+1.75(2.00)}_{-2.79(3.21)}$} & {$-0.019^{+1.59(1.74)}_{-2.01(2.70)}$} & {$-0.67^{+3.66(3.66)}_{-5.33(5.33)}$} & {$0.50^{+0.20(0.20)}_{-0.30(0.30)}$} \\ \hline
    {OHD+SN+BAO} & {$0.70^{+0.007(0.015)}_{-0.008(0.015)}$}  & {$0.98^{+1.73(1.77)}_{-1.94(2.04)}$} & {$0.016^{+1.57(1.76)}_{-1.42(1.67)}$} & {$-1.79^{+4.35(4.72)}_{-4.21(4.21)}$} & {$ 0.40^{+0.29(0.29)}_{-0.20(0.20)}$} \\ \hline\hline

    \end{tabular}
    \caption{Best-fit parameters of the Hubble rate, the function $G$ and its derivatives at $z=0$ and the equivalence redshift at 1$\sigma$-2$\sigma$ for the OHD+SN and OHD+SN+BAO data sets in the case of a non-flat universe.}
    \label{bestfitnonflat}
\end{table}

As stated above, in our analyses we made a set of approximations that clearly underestimate the real errors associated with our measurements. These approximations are related to the truncations we carried out throughout our study, in fact we faced

\begin{itemize}
    \item[-] approximation of spatial curvature to invert the relation implying the equivalence redshift in the case of a non-flat, with the aim of simplifying Eq. \eqref{zeqnonflat}, in order to get the expressions of the deceleration and jerk parameters;

    \item[-] truncation of the cosmographic terms, \textit{i.e.}, luminosity and angular distances,  Hubble rate, up  to the second order, prompting errors due to the fixed order we decide to truncate the series.
\end{itemize}

In view of the approximations made, we can thus think to increase the errors by

\begin{eqnarray}
    \sigma = \sqrt{\sum_{\nu}\sigma_\nu^2+\sum_{\mu}\sigma_\mu^2},
\end{eqnarray}

where $\sigma_\nu$ are the confidence levels shown in Tabs. \ref{bestfitflat}-\ref{bestfitnonflat}, whereas $\sigma_\mu$ are the errors induced by the previously mentioned approximations. In particular, we quantify the induced error bars by virtue of a maximum set up to $\lesssim 30\%$ of discrepancy. This value is reached numerically from our computations as one compare the real Hubble function with the approximated one. Specifically, the main discrepancies come from the approximations made for the non-flat case, where we related $\Omega_k$ to the matter and from truncations as above stressed.

The corresponding mean values of the measured coefficients, reported with bars on top, then read

\begin{subequations}
\begin{align}
{\rm \bf Flat\,\,case\,} && {\rm \bf Non-flat\,\,case\,} \nonumber \\
\bar h_0&= 0.69^{+0.10(0.11)}_{-0.10(0.11)} & \bar h_0&= 0.70^{+0.22(0.23)}_{-0.22(0.23)}\\
\bar G(z_{eq})&=0.98^{+2.33(3.07)}_{-1.18(1.32)}, & \bar G(z_{eq})&= 1.03^{+2.47(2.68)}_{-3.40(3.81)}\\
\bar G_0^\prime&=0.40^{+0.83(1.03)}_{-1.51(1.60)} & \bar G_0^\prime&= -0.0015^{+2.25(2.48)}_{-2.48(3.18)}\\
\bar G_0^{\prime\prime}&=0.04^{+1.47(1.47)}_{-0.74(0.74)} & \bar G_0^{\prime\prime}&= -1.23^{+5.69(5.98)}_{-6.79(6.79)}\\ \nonumber
\end{align}
\end{subequations}

while

\begin{align}
{\rm \bf Flat\,\,case\,} && {\rm \bf Non-flat\,\,case\,} \nonumber \\
\bar z_{eq}&= 0.51^{+0.29(0.29)}_{-0.45(0.45)} &
\bar z_{eq}&= 0.45^{+0.42(0.42)}_{-0.43(0.43)}
\end{align}

The errors inside the mean values are calculated considering the errors in Tabs. \ref{bestfitflat}-\ref{bestfitnonflat} together with the errors induced by the approximations we used throughout our work, $\sim 4\%$ and $\sim 9\%$ in the flat case for the OHD+SN and OHD+SN+BAO data set, respectively, while $\sim 12\%$ and $\sim 19\%$ in the non-flat case for the OHD+SN and OHD+SN+BAO data set, respectively.

\section{Evolution of the equation of state of the universe}\label{sezione5}

As stated above, the equation of state of the universe is simply $\omega=P/\rho$, where $P=\sum_i P_i$ is the total pressure of all constituents entering the second Friedmann equation, while $\rho=\sum_i \rho_i$ is the total density, composed by the sum of all constitutents entering the first and second Friedmann equations.

Following the same procedure above reported, we here are interested in evaluating $\omega$ for the flat and non-flat scenarios.

In the flat case, combining the r.h.s. of Eqs. \eqref{Fried} and \eqref{q_j_definition}, we have

\begin{eqnarray}\label{wflat}
    \omega=\frac{2q-1}{3},
\end{eqnarray}
valid for any redshift. At our time, the above relation reduces to $\omega_0=\frac{2q_0-1}{3}$,
where $q_0$ in the flat scenario is Eq. \eqref{q0zeq}. After plugging inside Eq. \eqref{wflat} the deceleration parameter in the flat case, at $z=0$, we get

\begin{eqnarray}\label{wflat1}
    \omega_0 = \frac{1}{3}\left[\frac{G(z_{eq})-(1+z_{eq})^3(2-G^\prime_0)}{G(z_{eq})+(1+z_{eq})^3}-1\right].
\end{eqnarray}

The variation of the equation of state of the universe depends on the derivative of $q$, as one can argue from Eq. \eqref{wflat}. Thus, since   \cite{reviewmia}

\begin{eqnarray}\label{qj}
    j(z) = (1+z)\frac{dq}{dz}+2q^2+q\,,
\end{eqnarray}
it is possible to write the first derivative of $\omega$ as function of the jerk parameter. To do so, we first differentiate with respect to $z$ Eq. \eqref{wflat}, getting  $\frac{d\omega}{dz}=\frac{2}{3}\frac{dq}{dz}$ and then, by virtue of Eq. \eqref{qj}, we infer

\begin{eqnarray}\label{derwflat}
 \omega^\prime\equiv   \frac{d\omega}{dz}= \frac{2}{3(1+z)}\left[j-2q^2-q\right],
\end{eqnarray}

reducing to $\omega^\prime_0=\frac{2}{3}\left[j_0-2q_0^2-q_0\right]$, and so, plugging Eq. \eqref{j0zeq} yields

\begin{multline}\label{derwflat1}
    \omega^\prime_0=\frac{2G(z_{eq})+(1+z_{eq})^3(2-2G^\prime_0-G^{\prime\prime}_0)}{3[G(z_{eq})+(1+z_{eq})^3]}-\frac{[G(z_{eq})-(1+z_{eq})^3(2-G^\prime_0)]^2}{3[G(z_{eq})+(1+z_{eq})^3]^2}-\frac{G(z_{eq})-(1+z_{eq})^3(2-G^\prime_0)}{3[G(z_{eq})+(1+z_{eq})^3]}.
\end{multline}

Analogously, $\omega$ in the non-flat scenario is given by

\begin{eqnarray}\label{wnonflat}
    \omega=\frac{1}{3(1-\Omega_k/a^2)}\left[2{q}-1+\frac{\Omega_k}{a^2}\right],
\end{eqnarray}

where we can  conventionally assume that at present time $a_0=1$ leading to $\omega_0=\frac{1}{3(1-\Omega_k)}\left[2q_0-1+\Omega_k\right]$. Thus, ${q}_0$ in a non-flat universe is given by Eq. \eqref{qtildenonflat} and we get

\begin{eqnarray}\label{wnonflat1}
     \omega_0=\frac{1}{3(1-\Omega_k)}\left[\frac{(1+z_{eq})^3(G^\prime_0-2)(1-\Omega_k)-G(z_{eq})(G^\prime_0-3)}{(1+z_{eq})^3}-1+\Omega_k\right].
\end{eqnarray}

Similarly, in a non-flat scenario, we differentiate Eq. \eqref{wnonflat} with respect to the redshift so that we have

\begin{eqnarray}
    \omega^\prime \equiv \frac{d\omega}{dz} = \frac{2\left(j-2q^2-q\right)}{3(1+z)\left[1-\Omega_k(1+z)^2\right]}+\frac{4q(1+z)\Omega_k}{3\left[1-\Omega_k(1+z)^2\right]^2}.
\end{eqnarray}

At present times

\begin{eqnarray}\label{derwnonflat}
    \omega^\prime_0=\frac{2}{3(1-\Omega_k)}\left[{j}_0-2{q}^2_0-{q}_0\right]+\frac{4q_0\Omega_k}{3(1-\Omega_k)^2},
\end{eqnarray}

where ${j}_0$ is given by Eq. \eqref{jtildenonflat}.

Substituting the deceleration and jerk parameters in the case of a non-flat universe inside Eq. \eqref{derwnonflat} yields

\begin{eqnarray}\label{derwnonflat1}
   \omega^\prime_0&=&\frac{2(1-G^\prime_0)+G^{\prime\prime}_0}{3(1-\Omega_k)}-\frac{\Omega_k(2-2G^\prime_0-G^{\prime\prime}_0)}{3(1-\Omega_k)}-\frac{G(z_{eq})(G^{\prime\prime}_0-2G^\prime_0)}{3(1+z_{eq})^3(1-\Omega_k)}-\frac{(1-\Omega_k)(G^\prime_0-2)^2}{3} \nonumber+\\&-&\frac{G^2(z_{eq})(G^\prime_0-3)^2}{3(1+z_{eq})^6(1-\Omega_k)}+\frac{2G(z_{eq})(G^\prime_0-2)(G^{\prime}_0-3)}{3(1+z_{eq})^3}-\frac{(G^\prime_0-2)}{3}+\frac{G(z_{eq})(G^\prime_0-3)}{3(1-\Omega_k)(1+z_{eq})^3}+\\&+&\frac{2\Omega_k(G^\prime_0-2)}{3(1-\Omega_k)}-\frac{2\Omega_kG(z_{eq})(G^\prime_0-3)}{3(1-\Omega_k)^2(1+z_{eq})^3}. \nonumber
\end{eqnarray}

Consequently, we substitute inside Eq. \eqref{wflat1} and Eq. \eqref{derwflat1} the values inside Tab. \ref{bestfitflat} for the OHD+SN and OHD+SN+BAO data set in the case of a flat universe and Eq. \eqref{wnonflat1} and Eq. \eqref{derwnonflat1} the values inside Tab. \ref{bestfitnonflat}  for the OHD+SN and OHD+SN+BAO data set in the case of a non-flat universe leading to the numerical values for $\omega_0$ and $\omega_0^\prime$. To this end, we compute the errors through the logarithmic chain rule, namely
\begin{equation}
\delta \theta_j=\sum_i\left|\frac{\partial \theta_j}{\partial x_i}\right|\delta x^i\,,
\end{equation}
where $\theta\equiv(\omega,\omega^\prime)$ and $x_j\equiv\left(G_0^\prime,G_0^{\prime\prime},G(z_{eq}),\Omega_k,z_{eq}\right)$. The corresponding numerical bounds are reported in Tab. \ref{wderwflatnonflat}.

\begin{table}[H]
    \centering
     \setlength{\tabcolsep}{1.em}
    \renewcommand{\arraystretch}{1.3}
    \begin{tabular}{cccc} \hline\hline
    {Cases} & {Data set} & {$\omega_0$} & {$\omega_0^\prime$}  \\ \hline
    \multirow{2}{*}{Flat} & {OHD+SN} & {$-0.70^{+0.45(0.61)}_{-0.59(0.61)}$} & {$0.36^{+0.79(1.01)}_{-0.81(0.84)}$}  \\
     {} & {OHD+SN+BAO} & {$-0.66^{+0.43(0.48)}_{-0.34(0.41)}$} & {$0.52^{+0.66(0.68)}_{-0.43(0.45)}$} \\ \hline\hline
    \multirow{2}{*}{Non-flat} & {OHD+SN} & {$-0.70^{+0.99(1.10)}_{-1.44(1.72)}$} & {$0.80^{+1.90(2.02)}_{-2.80(3.10)}$}  \\
     {} & {OHD+SN+BAO} & {$-0.66^{+0.79(0.84)}_{-0.71(0.78)}$} & {$1.19^{+2.10(2.23)}_{-2.05(2.14)}$} \\ \hline\hline

    \end{tabular}
    \caption{Values of $\omega$ and its derivative at 1$\sigma$-2$\sigma$ using the values of the function $G$ and its derivatives at $z=0$ and the equivalence redshift for the OHD+SN and OHD+SN+BAO data set for the flat and non-flat scenarios.}
    \label{wderwflatnonflat}
\end{table}

\subsection{Theoretical consequences on dark energy evolution}

In order to compare our findings, reported in Tab. \ref{wderwflatnonflat}, with current bounds over the dark energy evolution, let us consider the predictions made by the $\Lambda$CDM paradigm, where $G(z)=1$ at all times.

Specifically, our results show that acceptable mean values for $\omega_0$ and $\omega_0^\prime$ span as

\begin{itemize}
    \item[-] \emph{{\bf Flat case.}}
\begin{equation}
\bar \omega_0=-0.68^{+0.63(0.78)}_{-0.69(0.74)}, \quad {\rm and},\quad \bar \omega_0^\prime= 0.44^{+1.03(1.22)}_{-0.92(0.95)}\,,
\end{equation}
    \item[-] \emph{{\bf Non-flat case.}}
\begin{equation}
\bar \omega_0= -0.68^{+1.29(1.40)}_{-1.62(1.90)}, \quad {\rm and},\quad \bar \omega_0^\prime= 1.00^{+2.84(3.02)}_{-3.48(3.77)}\,,
\end{equation}
\end{itemize}

where, we assumed $\bar \omega_0$ and $\bar\omega_0^\prime$ as the mean values computed between the two measurements obtained in the flat and non-flat cases, respectively, while the total errors are computed adopting the upper and lower errors for each parameters by the relations  $\sqrt{\sigma_{upper}^2+\sigma_{upper}^2}$ and $\sqrt{\sigma_{lower}^2+\sigma_{lower}^2}$.

The corresponding bounds are close to the average values $\omega_0\sim -0.7$ and $\omega_0^\prime\sim 0.4$, while for the non-flat case stronger discrepancies set $\omega_0^\prime\sim 1$. These discrepancies occur as a consequence of the presence of $\Omega_k$ in the deceleration and jerk parameters. Indeed, the universe equation of state can be inferred from the second Friedmann equation that puts in correspondence $\omega$ with the deceleration parameter. However, as one invokes the jerk parameter, through Eq. \eqref{qj}, derivatives of $q$ appear, i.e., leading to derivatives of the spatial curvature, see Eq. \eqref{wnonflat}. This implies that the corresponding values may significantly change in the final output over $\omega^\prime$.

Hence, good agreement with the current cosmological model. Results from Tab. \ref{bestfitflat} and Tab. \ref{bestfitnonflat} show that for both data sets the value of the Hubble constant is in agreement with the one in Ref. \cite{planck}. Values of the function $G(z_{eq})$ are also in agreement with the assumption that $G(z_{eq})\gtrsim1$ so that $z_{eq}$ occurs in the past.

The equivalence redshift, in both scenarios of flat and non-flat universes is totally compatible with the value predicted by the $\Lambda$CDM model. All of this suggests that if there is a departure from the standard cosmological model it is extremely small.

Regarding the results we obtained for the equation of state and its derivative we argue that, even in this case our findings are compatible with the ones expected for the cosmological paradigm. In fact, we stress that here we are not considering the equation of state in which only the dark energy term is involved but we are considering the equation of state with the matter and dark energy terms.

In order to show that we consider the continuity equation

\begin{eqnarray}
    \frac{d\rho}{dt} = -3H(\rho+P).
\end{eqnarray}

We rewrite the previous expression considering $\omega = P/\rho$ and $dt = -\frac{dz}{(1+z)H}$. In so doing, we get

\begin{eqnarray}\label{conteq}
    \frac{d\rho}{dz} = \frac{3(1+\omega)}{(1+z)}\rho.
\end{eqnarray}

Now, we recast the energy density as $\rho = \rho_m+\rho_\Lambda=\rho_c[\Omega_m(1+z)^3+\Omega_\Lambda]$ where $\rho_c$ is the critical density. In this way, we get

\begin{eqnarray}
    \omega = -\frac{1}{1+\frac{\Omega_m}{1-\Omega_m}(1+z)^3},
\end{eqnarray}

and its derivative

\begin{eqnarray}
    \omega^\prime = \frac{3\Omega_m(1+z)^2}{(1-\Omega_m)\left[1+\frac{\Omega_m}{1-\Omega_m}(1+z)^3\right]^2}.
\end{eqnarray}

Substituting the matter density from Ref. \cite{planck} we get

\begin{eqnarray}
    \omega_0 = -0.68\pm 0.007 \quad {\rm and} \quad \omega^\prime_0 = 0.65\pm 0.008.
\end{eqnarray}

In the non-flat scenario the energy density is $\rho = \rho_m+\rho_\Lambda+\rho_k = \rho_c[\Omega_m(1+z)^3+\Omega_k(1+z)^2+\Omega_\Lambda]$. Thus, plugging $\rho$ in this scenario inside Eq. \eqref{conteq} yields

\begin{eqnarray}
    \omega = -\frac{1}{3}\frac{\Omega_k(1+z)^2+3(1-\Omega_m-\Omega_k)}{\Omega_m(1+z)^3+\Omega_k(1+z)^2+(1-\Omega_m-\Omega_k)},
\end{eqnarray}

and its derivative

\begin{eqnarray}
    \omega^\prime = \frac{1}{3}\frac{\Omega_k\Omega_m(1+z)^4+(1+z)(1-\Omega_m-\Omega_k)(9\Omega_m+4\Omega_k)}{[\Omega_m(1+z)^3+\Omega_k(1+z)^2+(1-\Omega_m-\Omega_k)]^2}.
\end{eqnarray}

Plugging the matter density and spatial curvature from the Planck collaboration leads to

\begin{eqnarray}
    \omega_0 = -0.72^{+0.026}_{-0.019}  \quad {\rm and} \quad \omega^\prime_0 = 0.64^{+0.018}_{-0.015}.
\end{eqnarray}

Confronting our results with the aforementioned values we argue that our findings are totally compatible at both $1\sigma-2\sigma$ with the cosmological paradigm. This suggests an overall agreement with a constant dark energy term even if it is not excluded that we might have a slight evolution of dark energy at higher redshifts.

\section{Final remarks and perspectives}\label{sezione6}

The determination of the transition and equivalence epochs between dark energy and matter can provide valuable insights into the nature of dark energy at redshifts beyond the present time, $z=0$. These quantities are particularly significant as they allow us to characterize the behavior of dark energy at earlier cosmic times. The transition epoch  marks the onset of cosmic acceleration and corresponds to the point where the deceleration parameter becomes zero. This fact makes it relatively easier to be found, compared to the equivalence between dark energy and dark matter, which is less commonly explored in the literature. This equivalence redshift, denoted as $z_{eq}$, plays a crucial role in understanding the interplay between dark energy and dark matter in the cosmic evolution.

In order to reconstruct the equivalence redshift, we have developed a cosmographic approach that establishes a relationship between fundamental kinematic coefficients at $z_{eq}$, removing the cosmic mass and healing the degeneracy problem. Specifically, to compute $q_0$ and $j_0$ at our current time, we replaced the mass with the one computed at the equivalence redshift. Hence, we showed that the corresponding cosmographic series  is well-suited for cosmographic  distances in order to evaluate bounds over $z_{eq}$ and on the dark energy evolution. Indeed, assuming a generic dark energy model in the form of perfect barotropic fluid, we fixed constraints on dark energy today, $G_0\equiv G(z=0)$, at the equivalence redshift $G(z_{eq})$, involving variations of these quantities expressed in terms of first derivatives computed at our times. This permitted us to write up in a model-independent way both $q_0$ and $j_0$ in terms of the redshift at which the  equivalence itself occurs.

Our numerical analyses were performed using a MCMC code written in Python, making use of the Metropolis-Hastings algorithm. We performed a twofold analysis. First, we investigated whether the effects induced by spatial curvature are relevant or not in the overall numerical procedure. Second, we investigated plausible constraints on the equivalence between baryons and dark energy, then on the equivalence between cold dark matter and dark energy and finally on total mass and dark energy.

To work this out, we expanded around our time the luminosity and angular distances, together with a direct Hubble rate expansion. Afterwards, we  considered the most recent Pantheon SNe Ia, OHD, and BAO data sets. Finally, we directly fitted our expanded distances, adopting a hierarchy among data sets, namely we first focused on SNe Ia and OHD and secondly we included the BAO data points. In such a way, by binning the unknown dark energy function within precise ranges of priors, we emphasized how bounds over parameters changed as more data points are included into computation. In this respect, we compared our findings with those obtained in three cosmological models, namely the $\Lambda$CDM concordance background, the $\omega$CDM scenario and the CPL parametrization. Thus, we discussed the consequences of our numerical limits on the equation of state of the universe and on its variation.

In so doing, our error bars have been also computed including the systematic effects induced by all the approximations here adopted, mainly due to  series truncation and negligible higher-order terms in $\Omega_k$. We compared our findings with the theoretical expectations obtained from the aforementioned three paradigms. Hence, our scenarios clearly showed that, albeit the best background model is still provided by the $\Lambda$CDM paradigm, the limits on dark energy evolution cannot exclude a slight evolution at intermediate redshifts. Consequently, it is not possible to exclude the $\omega$CDM model, albeit the CPL parametrization showed less suitable numerical results.

In view of our results, we expect to improve our fits by adding new data catalogs and checking whether this can modify the results on the dark energy variation. We will also explore possible additional equivalence and/or transition redshifts among dark energy and other constituents of the universe, extending the work \cite{luongomuccinogrb}.

\section*{Acknowledgements}
The authors acknowledge Marco Muccino for fruitful discussions on the topic related to the numerical analysis of cosmic data. CC acknowledges the hospitality of the United States Air Force Research Laboratory (AFRL) in Rome-NY where part of  this work was carried out. SC acknowledges the Istituto Nazionale di Fisica Nucleare (INFN), Sezione di Napoli, {\it iniziativa specifica} QGSKY. The work of OL is  partially financed by the Ministry of Education and Science of the Republic of Kazakhstan, Grant: IRN AP19680128. This paper is partially based upon work from the COST Action CA21136, \textit{Addressing observational tensions in cosmology with systematics and fundamental physics} (CosmoVerse) supported by COST (European Cooperation in Science and Technology).

\clearpage

\begin{appendices}

\section{Contour plots}

We here report the contour plots obtained by numerical analyses. They are evaluated using the Metropolis-Hastings algorithm. The prompt on their right  corresponds to the likelihood shapes with 1$\sigma$-2$\sigma$ confidence levels. The different colors  correspond to the $68\%$ and $95\%$ of probability.

\begin{figure}[H]
    \centering
    \includegraphics{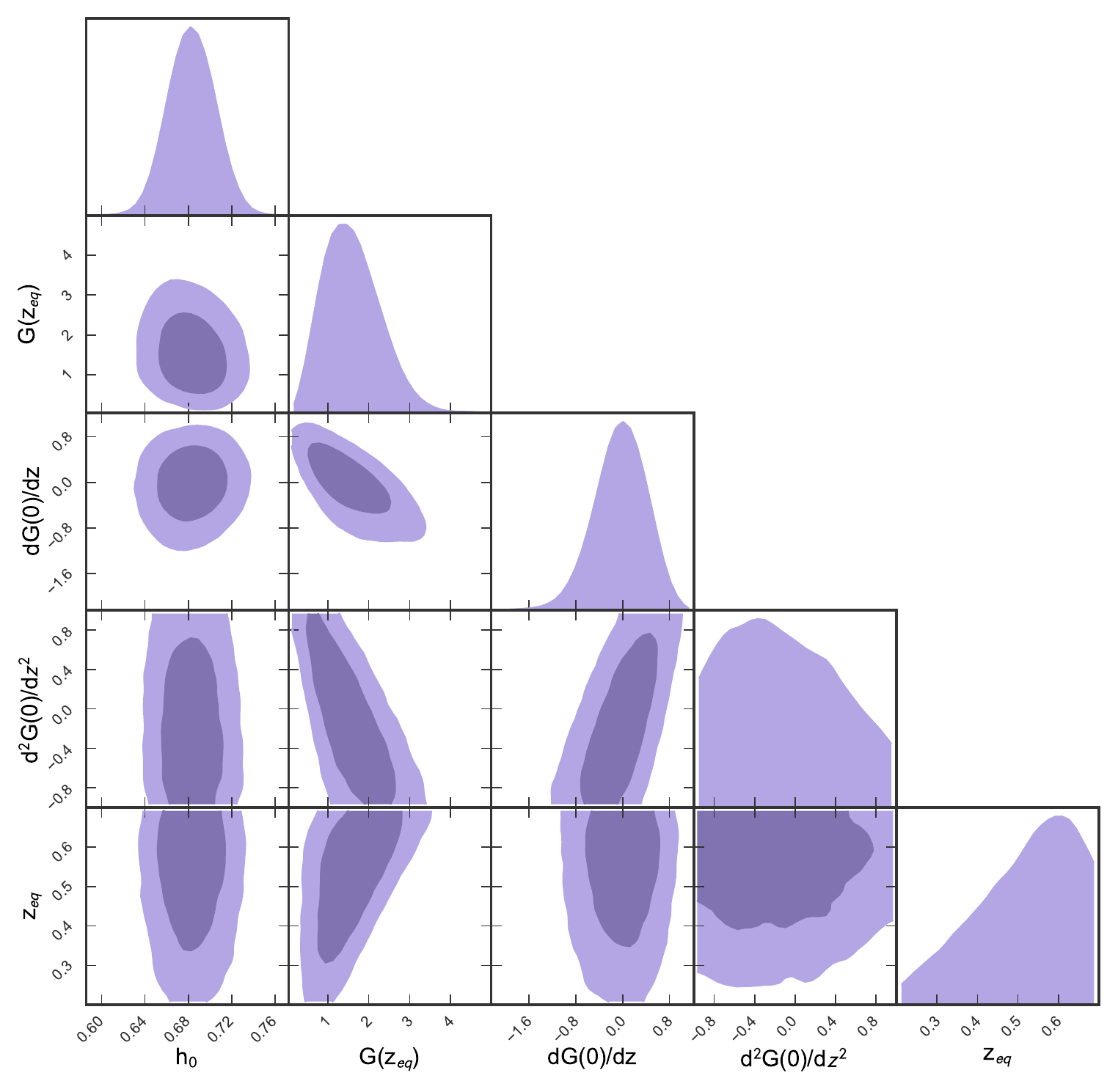}
    \caption{Contour plot of the best-fit parameters for the OHD+SN data set of the Hubble rate, the function $G$ and its derivatives at $z=0$ and the equivalence redshift in the case of a flat universe. Dark purple area marks the 1$\sigma$ confidence regions while light purple area marks the 2$\sigma$ confidence regions.}
    \label{contourOHDSN}
\end{figure}

\begin{figure}[H]
    \centering
    \includegraphics{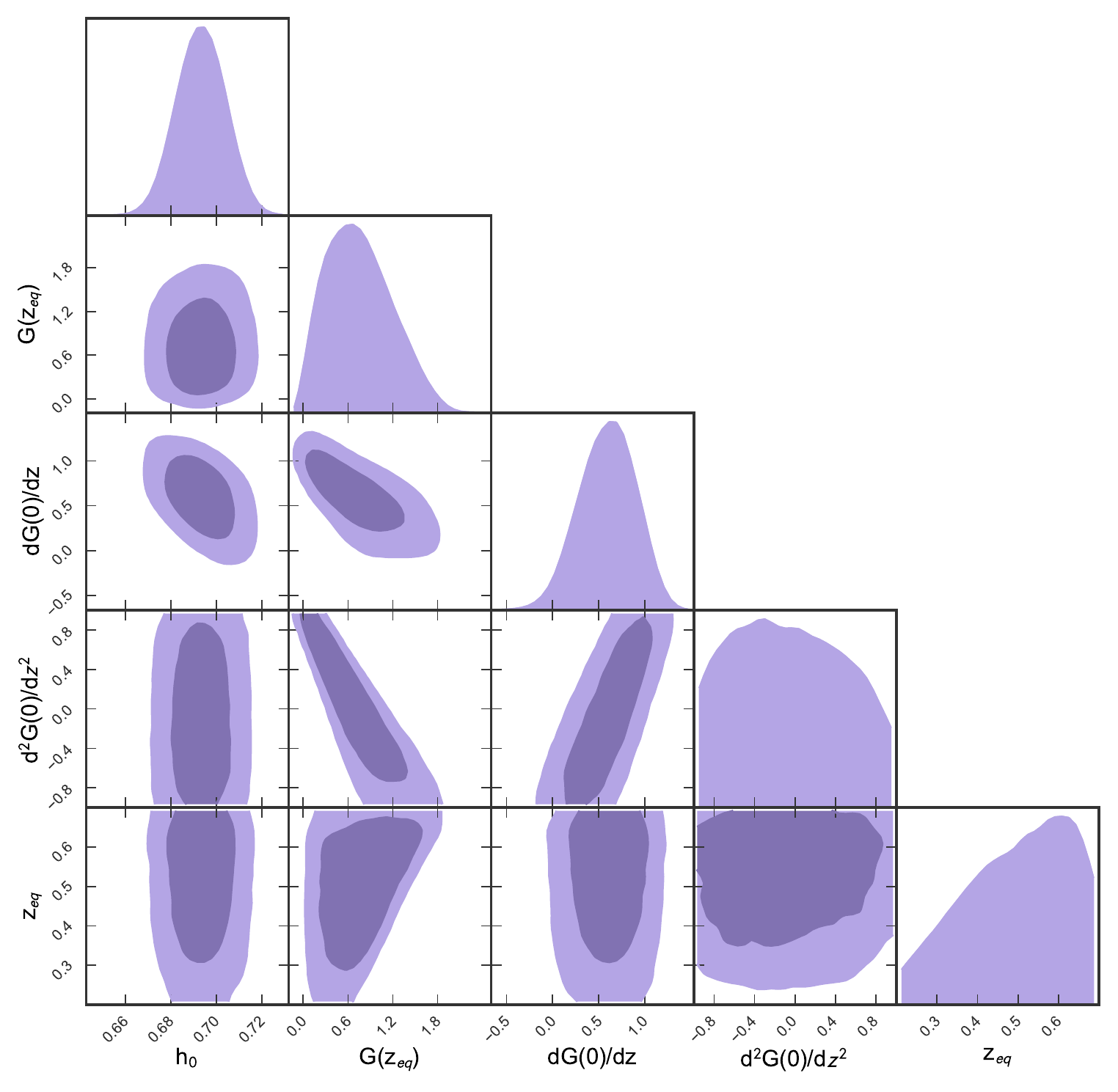}
    \caption{Contour plot of the best-fit parameters for the OHD+SN+BAO data set of the Hubble rate, the function $G$ and its derivatives at $z=0$ and the equivalence redshift in the case of a flat universe. Dark purple area marks the 1$\sigma$ confidence regions while light purple area marks the 2$\sigma$ confidence regions.}
    \label{contourOHDSNBAO}
\end{figure}

\begin{figure}[H]
    \centering
    \includegraphics{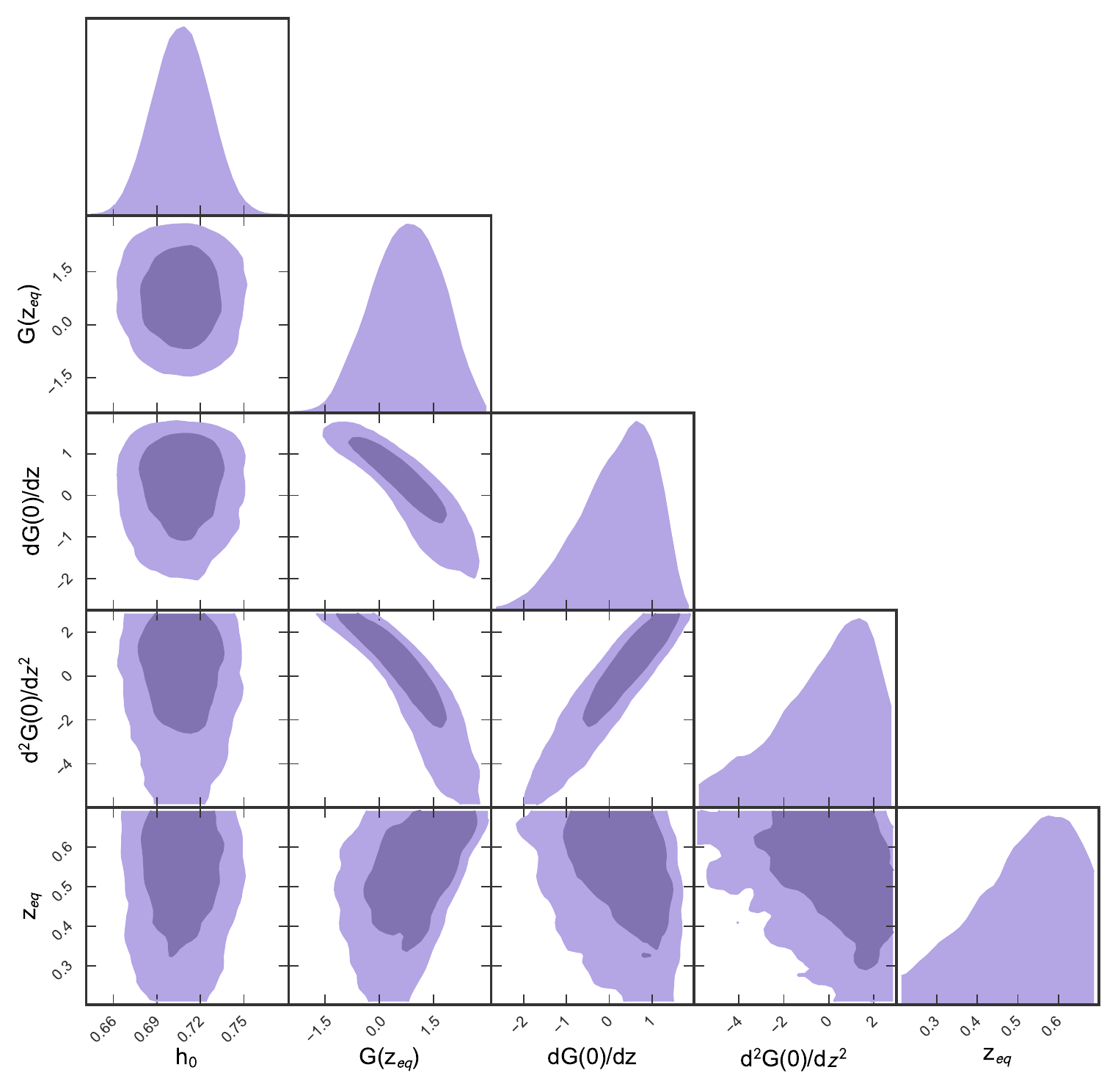}
    \caption{Contour plot of the best-fit parameters for the OHD+SN data set of the Hubble rate, the function $G$ and its derivatives at $z=0$ and the equivalence redshift in the case of a non-flat universe. Dark purple area marks the 1$\sigma$ confidence regions while light purple area marks the 2$\sigma$ confidence regions.}
    \label{contourOHDSNcurvature}
\end{figure}

\begin{figure}[H]
    \centering
    \includegraphics{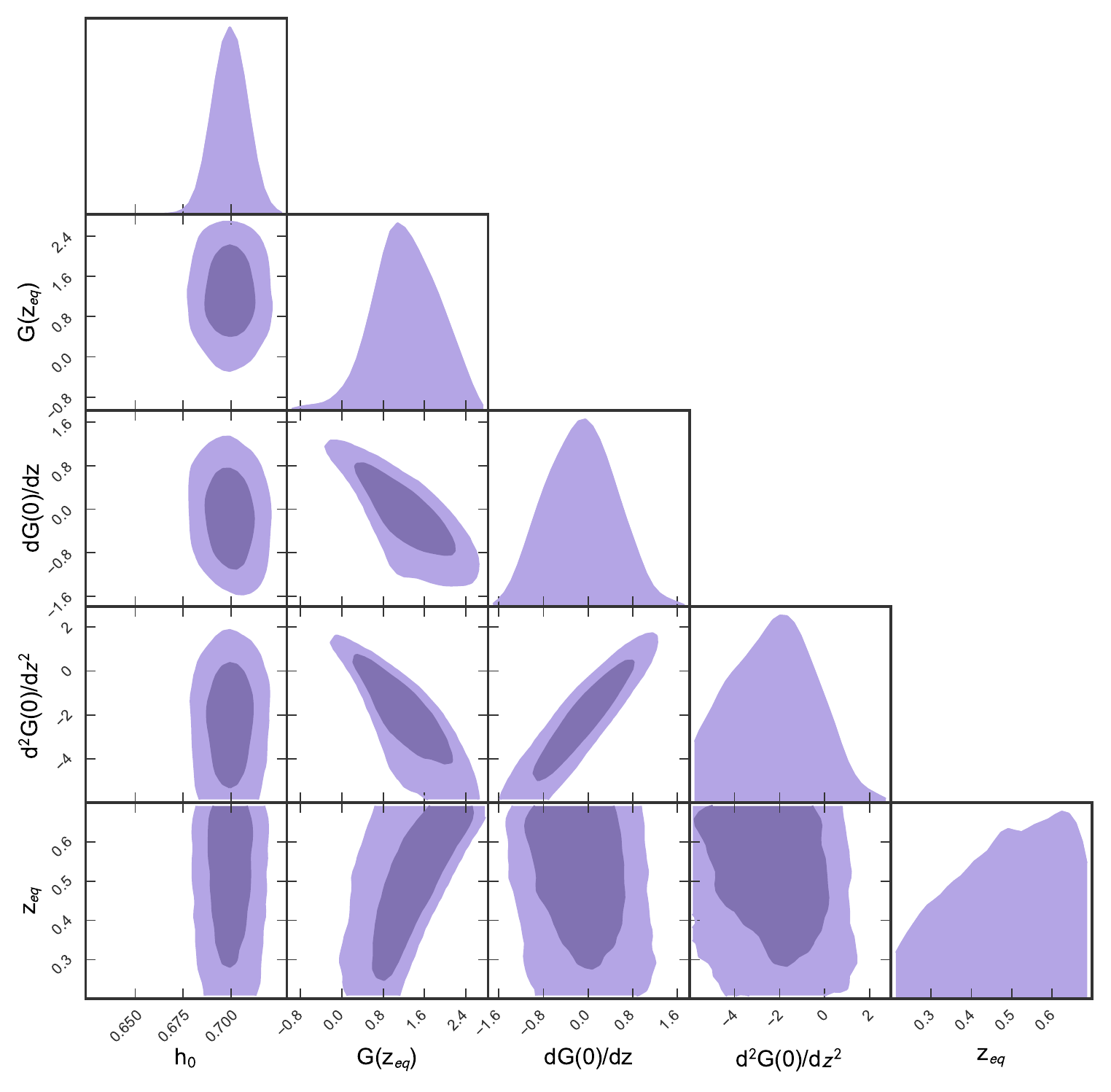}
    \caption{Contour plot of the best-fit parameters for the OHD+SN+BAO data set of the Hubble rate, the function $G$ and its derivatives at $z=0$ and the equivalence redshift in the case of a non-flat universe. Dark purple area marks the 1$\sigma$ confidence regions while light purple area marks the 2$\sigma$ confidence regions.}
    \label{contourOHDSNBAOcurvature}
\end{figure}

\section{The equivalence redshift for $\Omega_k \neq 0$}\label{zeqOkdiv0}

As reported in the text, the whole equation involving $\Omega_k$ needs simplifications in order to get a quasi-analytical solution from Eq. \eqref{zeqnonflat}. Specifically, we suggest how to handle $\Omega_k$ by considering that approximately $\Omega_m \sim 0.3$  and $\Omega_k \sim -0.05$, following the Planck results. In view of the fact that $\Omega_k$ is unequivocally smaller than matter and dark energy densities, we can conventionally write $\Omega_k = x\Omega_m \rightarrow x = -\frac{\Omega_k}{\Omega_m} = \frac{1}{6}$, yielding $\Omega_k\approx-\frac{\Omega_m}{6}$. This mathematical trick enables us to simply the mathematical complexity of our underlying equations and to get suitable and quite compact solutions. Clearly, this ansatz does not alter the physics of our problem, leaving unaltered the results that we prompted. So, using this ansatz in Eq. \eqref{zeqnonflat} we get

\begin{eqnarray}\label{zeqcalc}
    z_{eq} = -1+\frac{1}{6\Omega_m}\left[\frac{2^{\frac{4}{3}}\Omega_m^2}{36y(\Omega_m)}+2^{\frac{2}{3}}y(\Omega_m)+\frac{\Omega_m}{3}\right],
\end{eqnarray}

where

\begin{align}\label{zeqcalc1}
    y(\Omega_m) = \left[\frac{\Omega_m^3}{108}-\frac{27}{6}G(z_{eq})\Omega_m^2(5\Omega_m-6)+3^{\frac{3}{2}}\sqrt{\frac{G(z_{eq})}{6}\Omega_m^2(5\Omega_m-6)\left(-\frac{\Omega_m^3}{54}+\frac{27}{6}G(z_{eq})\Omega_m^2(5\Omega_m+6)\right)}\right]^{\frac{1}{3}}.
\end{align}

We further simplify Eq. \eqref{zeqcalc} substituting inside Eqs. \eqref{zeqcalc} - \eqref{zeqcalc1} $\Omega_m \sim 0.3$ and $G(z_{eq}) \sim 1$ leading to

\begin{eqnarray}\label{zeqapprox}
    z_{eq} \approx -1+\frac{1}{6\Omega_m}\left[2^{\frac{2}{3}}y(\Omega_m)\right],
\end{eqnarray}

with
    $y(\Omega_m) \approx 2^{\frac{1}{3}}\left[27G(z_{eq})\Omega_m^2\right]^{\frac{1}{3}}$.

Afterwards, we substitute inside Eq. \eqref{zeqapprox} $\Omega_m$ written in terms of the deceleration parameter

\begin{eqnarray}
    \Omega_m = \frac{G^\prime_0-2(q_0+1)-\Omega_k(G^\prime_0-2)}{G^\prime_0-3}.
\end{eqnarray}

In this way, we get Eq. \eqref{qtildenonflat}.

Then, we plug in Eq. \eqref{zeqapprox} $\Omega_m$ written in terms of the jerk parameter

\begin{eqnarray}
    \Omega_m = \frac{2(j_0-1)-2G^\prime_0+G^{\prime\prime}_0-\Omega_k(2-2G^\prime_0-G^{\prime\prime}_0)}{G^{\prime\prime}_0-2G^\prime_0}.
\end{eqnarray}

In this way, we get Eq. \eqref{jtildenonflat}

\section{The equivalence redshift of baryons, cold dark matter and dark energy with $\Omega_k \neq 0$}\label{zeqOkdiv01}

First, in the case of baryons we have

\begin{eqnarray}\label{baryon}
    \Omega_{DE}G(z_{eq})&=\Omega_b(1+z_{eq,b})^3+\Omega_k(1+z_{eq,b})^2,
\end{eqnarray}

where $\Omega_b$ is the baryon density. In order to get $z_{eq,b}$ we first subtract Eq. \eqref{baryon} with Eq. \eqref{formal} so that we do not have the dark energy term. This leads to a third-grade equation whose solution is

\begin{eqnarray}
     z_{eq,b} = -\frac{b}{3a}+\sqrt[3]{-\frac{q}{2}+\sqrt{\frac{q^2}{4}+\frac{p^3}{27}}}+\sqrt[3]{-\frac{q}{2}-\sqrt{\frac{q^2}{4}+\frac{p^3}{27}}},
\end{eqnarray}

where

\begin{align}
    \frac{b}{3a} = 1+\frac{\Omega_k}{3\Omega_b}\,,\,\,\,\,
    q = \frac{2\Omega_k^3}{27\Omega_b^3}-\frac{(1+z_{eq})^2[\Omega_k+\Omega_m(1+z_{eq})]}{\Omega_b}\,,\,\,\,\,
    p = -\frac{\Omega_k^2}{3\Omega_b^2}.
\end{align}

Equivalently, for the cold dark matter case, we have

\begin{eqnarray}\label{baryoncdm}
    \Omega_{DE}G(z_{eq})=\Omega_{cdm}(1+z_{eq,cdm})^3+\Omega_k(1+z_{eq,cdm})^2
\end{eqnarray}

In the same fashion as done for the baryon case, we subtract Eq. \eqref{baryoncdm} with Eq. \eqref{formal} leading to a third-grade equation whose solution is

\begin{eqnarray}
     z_{eq,cdm} = -\frac{b}{3a}+\sqrt[3]{-\frac{q}{2}+\sqrt{\frac{q^2}{4}+\frac{p^3}{27}}}+\sqrt[3]{-\frac{q}{2}-\sqrt{\frac{q^2}{4}+\frac{p^3}{27}}},
\end{eqnarray}

where

\begin{align}
    \frac{b}{3a} = 1+\frac{\Omega_k}{3\Omega_{cdm}}\,,\,\,
    q = \frac{2\Omega_k^3}{27\Omega_{cdm}^3}-\frac{(1+z_{eq})^2[\Omega_k+\Omega_m(1+z_{eq})]}{\Omega_{cdm}}\,,\,\,
    p = -\frac{\Omega_k^2}{3\Omega_{cdm}^2}.
\end{align}

\end{appendices}

\end{document}